\documentclass[a4paper,fleqn]{cas-dc}
\usepackage[numbers,sort&compress]{natbib}

\usepackage{subcaption}
\usepackage{cleveref}
\usepackage{diagbox}
\usepackage{ragged2e} 
\usepackage{xurl}

\usepackage[modulo,switch]{lineno}   
\usepackage{amsmath}  
\usepackage{etoolbox} 

\newcommand*\linenomathpatch[1]{%
	\cspreto{#1}{\linenomath}%
	\cspreto{#1*}{\linenomath}%
	\csappto{end#1}{\endlinenomath}%
	\csappto{end#1*}{\endlinenomath}%
}

\linenomathpatch{equation}
\linenomathpatch{gather}
\linenomathpatch{multline}
\linenomathpatch{align}
\linenomathpatch{alignat}
\linenomathpatch{flalign}

\begin{document}

\shorttitle{Controlling anisotropy in 2D microscopic models of growth
}
\shortauthors{L.~Gagliardi et~al. }

\title [mode = title]{Controlling anisotropy in 2D microscopic models of growth
}

\author[1]{Luca Gagliardi}[orcid=0000-0003-3360-2537]

\cormark[1]
\ead{luca.gagliardi@iit.it}

\address[1]{Istituto Italiano di Tecnologia (IIT)-- Via Enrico Melen, 83
16152 Genoa
Italy}

\author[2]{Olivier Pierre-Louis}
[orcid = 0000-0003-4855-4822]
\ead{olivier.pierre-louis@univ-lyon1.fr}

\address[2]{Institut Lumière Matière -- UMR5306 CNRS
Université Lyon 1
Campus LyonTech - La Doua
10 rue Ada Byron
69622 Villeurbanne CEDEX, France }

\cortext[cor1]{Corresponding author}

\begin{abstract}
The quantitative knowledge of interface anisotropy in lattice models is a major issue,
both for the parametrization of continuum interface models, 
and for the analysis of experimental observations. 
In this paper, we focus on the anisotropy of line tension and stiffness,
which plays a major role both in equilibrium 
shapes and fluctuations, and in the selection of nonequilibrium growth patterns.
We consider a 2D Ising Hamiltonian on a square lattice with first and second-nearest-neighbor interactions.
The surface stiffness and line tension are calculated by means
of a broken-bond model for arbitrary orientations.
The analysis of the interface energy  allows us to determine the conditions under which
stiffness anisotropy is minimal.
These results are supported by a quantitative comparison with kinetic Monte Carlo simulations, 
based on the coupling of a field of mobile atoms
to a condensed phase.
Furthermore, we introduce a generic smoothing parameter which allows one to 
mimic the finite resolution of experimental microscopy techniques.
Our results provide a method to fine-tune the interface energy in models of nanoscale non-equilibrium processes, where anisotropy and fluctuations combine and give rise to non-trivial morphologies.

\end{abstract}


\maketitle

\section{Introduction}

Many growth processes, such as crystal growth from solution, electrochemical deposition, solidification, or viscous fingering in Hele-Shaw cells share similar mechanisms controlling pattern formation.
In particular, the surface tension (or line tension in 2D systems) and the stiffness are important control parameters in the selection of growth patterns \cite{Ben-Jacob1990,Saito1987,Saito1989,Saito1994,Uwaha1992,Plapp1997}.
Furthermore, surface tension anisotropy controls equilibrium shapes via classical Wulff construction \cite{Einstein2015,Saito1996}, and equilibrium fluctuations\cite{Misbah2010}.

Over the past decades, several strategies have been developed in order to model
the non-equilibrium dynamics of interfaces during growth processes.
In continuum models such as phase-field models or sharp interface models,
the surface tension with its full orientation-dependence
is given as an input parameter. 
As a consequence, continuum models can model any anisotropy.
However, the consistent inclusion of thermal fluctuations in these models (see, e.g. \cite{Gouyet2003})
leads to several difficulties, both in the quantitative validation 
of noise terms constrained by the fluctuation-dissipation theorem in non-equilibrium processes \cite{Misbah2010},
and in the design of efficient numerical schemes.
This is an important issue since thermal fluctuations play a major role 
for instance in the formation of side branches in dendrites \cite{Gouyet2003,Saito1987}.
Microscopic lattice-based approaches such as in kinetic Monte Carlo (KMC) simulations 
overcome this difficulty since they incorporate statistical fluctuations in a
self-consistent way, which is constrained by detailed balance at equilibrium, 
and which is still well defined far from equilibrium. 
However, the anisotropy of macroscopic quantities such as 
surface tension cannot be set arbitrarily in lattice models,
they emerge from the choice of microscopic model parameters, such as the bond
energy. 
Achieving a fine control of isotropy in lattice models is important for comparison to continuum growth models, 
where anisotropy acts as a singular perturbation~\cite{Saito1996}. 
In the literature, the introduction of next-nearest-neighbor interactions in lattice models has	been proposed as a route towards an increased control of interface anisotropy \cite{Plapp1997,Pierre-Louis2000,Einstein2007,Chame2020,Einstein2004}.


In this article, we aim at a quantitative description of the anisotropy of the surface tension and stiffness of
one-dimensional interfaces in two-dimensional square-lattice models with nearest and next-nearest-neighbor interactions.
An analytical expression for the orientation dependent line tension and stiffness was first derived in Ref.~\cite{Einstein2004}. 
However, a quantitative measure of the global degree of anisotropy of surface (or line) tension and stiffness as a function of such microscopic interactions and temperature, has not been achieved yet.
Our strategy is based on a quantitative comparison between analytical
predictions and kinetic Monte Carlo simulations. 
Our analytical results provide an expression of the surface tension and stiffness 
for arbitrary orientations
as a function of the nearest and next-nearest-neighbor interactions and of the temperature. 
In parallel, we present a KMC model based on the coupled dynamics of a dense ordered phase and a diffusing cloud of adatoms. Simulations support our analytical results at equilibrium and allow one to study efficiently nonequilibrium growth patterns.
We also show how to account for the finite resolution of experimental apparatus when measuring interface fluctuations. 
We mimic the finite resolution of experiments in our lattice simulations by performing a Gaussian convolution with a given smoothing parameter.
Finally, contrary to intuition, within the solid-on-solid description of the interface, the highest level of stiffness isotropy 
is obtained at finite temperatures. However the question of the existence of this isotropy optimum beyond
the solid-on-solid approximation is still open.

The paper is organized in the following way. We start with a presentation
of the kinetic Monte Carlo model and its numerical implementation. 
We then report the analytical calculation of the line tension and stiffness 
as functions of the step orientation in the presence of nearest-neighbor and next-nearest-neighbor interactions.
Then, we discuss the comparison at equilibrium between the analytic predictions and the KMC simulations. 
To conclude, we summarize our results and provide examples of nonequilibrium growth regimes.

\section{Kinetic Monte Carlo model}

We use a standard kinetic Monte Carlo algorithm \cite{Kotrla1996} with two types of atoms. 
This model accounts for a two-phase system, where a dense ordered phase,
for example a monolayer cluster of atoms or molecules on a flat substrate 
is in contact with a phase of non-interacting mobile units, that accounts for a concentration of diffusing atoms
or molecules on the substrate. At equilibrium, and at low enough temperatures, 
this system is equivalent to a 2D Ising square lattice \cite{Saito1996,Zandvliet2006}.
The deviations from low-temperature and low-density approximations
that can be observed in lattice models
due to the interactions between mobile atoms in the low-density phase (see, e.g.\cite{Krishnamachari1996})
are avoided here due to the absence of mutual interactions.
A similar two-phase KMC model, but with nearest-neighbors only, has already been reported in Ref.~\cite{Saito1989}.

In the following, we call the mobile units adatoms. Given the lattice constant $a$, each lattice site $x = i\times a $, $i=1, \dots, N$ with $N=L/a$, can be occupied by a solid
atom, by an adatom, or can be empty. Multiple occupancy of the same lattice site
is allowed only for adatoms, which do not interact among themselves nor with the solid bulk (i.e.~away from the interface). 
Adatoms diffuse in an isotropic way performing a random walk via hops to nearest-neighbors.  
Diffusion anisotropy was not considered in this work but can have important effects on the type of growth patterns observed \cite{Curiotto2019,Plapp1997,Danker2004} and would be a further source of shape selection in addition to interface energy effects which are the main focus of this paper.

The ordered phase is characterized by atoms with bonds along first and second-nearest-neighbor directions.
The bond energy to first-nearest-neighbors (NN) and second-nearest-neighbors (NNN) are 
$J_1$ and $J_2$ respectively. These energies are around $1$eV or a fraction of eV in atomic solids \cite{Steimer2001,Einstein2004,Misbah2010}.
We define the dimensionless bond energy ratio $\zeta = J_2/J_1$.  
The value of $\zeta$ controls the anisotropy of the 
interface between the two phases.
We do not consider the inclusion of trio (three atoms non pairwise) interactions. However in general, trio interactions account for only a small correction to the stiffness measured in experiments \cite{Einstein2004}.

The solid can grow or shrink at the expense of the adatom gas. 
Attachment and detachment processes occur at the edge of the solid phase.
Adatoms can attach to the solid (i.e.~can be transformed into a solid atom) when they have at least one NN  or NNN
in contact with the solid. Atoms of the solid that have at least one NN or NNN that is not in contact with the other atoms of the solid can detach and are then transformed into adatoms.

The kinetic Monte Carlo model is characterized by the rates of the possible microscopic events.
We use standard Arrhenius activation laws with an attempt frequency $\nu_0$.
The adatom hopping rate is 
\begin{align}
\label{eq:d_def}
d =\nu_0 \exp  [ -E_D/(k_BT) ],
\end{align}
where $k_BT$ is the thermal energy  and $E_D$ is the diffusion activation energy.
The attachment rate  reads 
\begin{align}
\label{eq:q_def}
q = \nu_0 \exp [ -E_Q/(k_BT) ],
\end{align}
with $E_Q$ the attachment/detachment activation energy.
Finally, the detachment rate at a given lattice site $i$ is 
\begin{align}
\label{eq:r_def}
r = \nu_0 \exp [-(J_1 (nn_i + \zeta nn'_i) + E_Q - E_S)/(k_BT)],
\end{align}
where $nn_i$ and $nn'_i$ are the number of nearest and next-nearest neighbors
around the site $i$.
The additional energy $E_S$ is an independent parameter that 
reduces the difference of energy between solid atoms and adatoms
so as to increase the adatom concentration when $E_S>0$.

\subsection{Equilibrium concentration}

Since the adatom diffusion rate from one site to its neighbor is
equal to the rate of the reverse move, detailed balance implies that equilibrium
within the adatom phase corresponds to a constant concentration. 
In global equilibrium of the system, composed of the adatom phase
and the solid phase, the adatom concentration reaches its equilibrium value.
This equilibrium concentration can be inferred from detailed balance at
the interface between the two phases. In equilibrium, 
the probability of a given configuration of the interface, denoted as $h$, is 
\hbox{$\pi(h) = \exp [-E(h)/(k_BT)]/Z$,} where $Z$ is the partition function.
In the presence of an equilibrium concentration $c_{eq}$, 
the probability for an adatom to be in a given site is $a^2c_{eq}$,
and the attachment rate reads $qa^2c_{eq}$.
Detailed balance \cite{Kotrla1996,Saito1996,Barabasi1995} implies
that the attachment rate  and the detachment rate $r$ are related via
\begin{equation}
	\label{eq:detailed_b}
\frac{qa^2c_{eq}}{r} = \frac{\pi(h)}{\pi(h-1)}= {\mathrm e}^{-\frac{E(h)-E(h-1)}{k_BT}}\, ,
\end{equation}
where $(h-1)$ is a notation that indicates an interface configuration that differs only by the detachment of one atom from the previous 
configuration.

The energy of the edge of the solid phase can be determined
from the evaluation of the number of broken bonds along the edge.
Since the breaking of one bond creates two broken bonds,
each broken bond to NN or NNN costs an energy $J_1/2$ or $J_2/2$ respectively.
The variation of the interface energy when removing
one atom from the interface is evaluated from the change of the number of broken bonds as
\begin{equation}
\label{eq:Delta_E_detach}
\begin{split}
 E(h) -E(h-1) &= J_1[2-nn_i + \zeta (2-nn'_i) ] \\
&= E_k -J_1(nn_i + \zeta nn'_i) \, ,
\end{split}
\end{equation}
with $E_k = 2J_1(1+ \zeta)$ the bond-breaking energy for the detachment
of a solid atom from a kink site.
Using \cref{eq:detailed_b} the equilibrium concentration therefore reads 
\begin{equation}
\label{eq:eq_conc}
c_{eq} = a^{-2}\frac{r}{q}{\mathrm e}^{-\frac{E(h) -E(h-1)}{k_BT}} 
=  a^{-2}{\mathrm e}^{-\frac{E_k-E_S}{k_BT}}\, .
\end{equation}
Two remarks are in order.
First, the possibility of obtaining 
a constant equilibrium concentration (that does not depend
on NN and NNN) results from the fact that
the dependence of the activation energy of $r$, in \cref{eq:r_def},
and of $E(h)-E(h-1)$, in \cref{eq:Delta_E_detach},
on  $nn_i$ and $nn'_i$ are identical. This is actually a constraint that
motivates the specific dependence on $nn_i$ and $nn'_i$ of the activation energy
of $r$ in \cref{eq:r_def}. 
As a second remark, the expression \cref{eq:eq_conc} confirms the role of the parameter $E_S$,
which tunes the equilibrium adatom concentration
without affecting the energy of the interface itself.

\subsection{Implementation details }
Our simulations are performed in a square lattice of size $L\times L$ with lattice constant $a$ and periodic boundary conditions.  
We employ a rejection-free BKL algorithm where events are grouped into classes \cite{Maksym1988,Kotrla1996}. 
The concentration of adatoms is set at the beginning of the integration
and the total number of atoms is therefore fixed during the dynamics.
Instead of the usual individual adatom diffusion events, we use collective diffusion events. 
In each collective diffusion event, all adatoms diffuse simultaneously.
Such collective diffusion event provides a speed-up of a factor seven with respect to the standard randomized single adatom move\footnote{The speed-up is not only due to avoiding the generation of $(n-1)$ random numbers, with $n$ the number of adatoms (in the standard implementation to move all adatoms we would need  $2\times n$ random numbers to select the diffusion event class \emph{and} the diffusion direction), but also to a more efficient memory access. Since lists are used to store events and random access is not possible in lists, looping through all lists elements at once is more efficient than accessing a given element.}.
Moreover, this move can be shared-memory parallelized, leading to a further performance increase, of approximately $10\%$ 
(for simulation box sizes larger than about $400\times 400$ or large concentrations).
Note that the possibility to simply cast diffusion in a simultaneous event is allowed by the absence of interactions between adatoms.

We choose to normalize time with the inverse of the adatom hopping rate $d^{-1}$,
length scales with the lattice spacing $a$, and energies with the NN bond
energy $J_1$. 
Normalized quantities are denoted with an overline.
Our simulations are therefore controlled by 4 dimensionless parameters:
$\overline{k_BT} = k_BT/J_1$,
$\zeta = J_{2}/J_1$,
$\bar{A} = (E_Q-E_D)/J_1$ and
$\bar{E}_S = E_S/J_1$.
All simulation results below are reported in these normalized units.

\section{Orientation-dependence of line tension and stiffness}

We here  derive general analytic expressions at equilibrium for the line tension, $\gamma(\theta)$, and stiffness, $\tilde{\gamma}(\theta) = \gamma(\theta) + \gamma''(\theta)$, 
where $\theta$ is the angle between the (10) direction and the average interface orientation.
Our main goal in the following is to provide
a detailed picture of the anisotropy of line tension and stiffness at finite temperatures.
As a preamble, we start with a short discussion
of the zero-temperature case.

\subsection{A simple picture at zero temperature}


At zero temperature  the free energy reduces to 
the energy, and it is sufficient to resort to a bond-counting strategy, 
where each nearest-neighbor or next-nearest neighbor broken 
bond contributes with an energy $J_1/2$ or $J_2/2$, respectively. Assuming 
a one-dimensional interface with regularly-spaced kinks separated by $m\geq 1$ sites along the (10) direction,
we may count the number of broken bonds. Each atom
on a straight (10) edge exhibits one NN broken bond and two NNN broken bonds.
Each kink leads to an additional NN in the energy per unit length.
Dividing the broken bond energy between two kinks by the distance between two kinks, 
we obtain the broken bond energy per unit length of the interface
\begin{align}
\bar\epsilon_m= \frac{m(\frac{1}{2}+\zeta)+\frac{1}{2}}{(m^2+1)^{1/2}}\,.
\end{align}
The integer $m$ can be related to the orientation of the interface via
the relation $m=\mathrm{cot}|\theta|$, where $\theta$ is the angle of the interface
with the (10) orientation, with $0\leq|\theta|\leq\pi/4$. The minimum of $\bar\epsilon_m$ is reached for
$m\rightarrow +\infty$ corresponding to the (10) orientation with $\theta=0$ for $\zeta<2^{-1/2}$,
and for $m=1$ corresponding to the (11) orientation with $\theta=\pi/4$ when $\zeta>2^{-1/2}$.
As a consequence, the two limits of small $\zeta$ and large $\zeta$ lead to square low temperatures 
equilibrium shapes with sides along the (10) or (11) orientation, respectively. 
When $\zeta=2^{-1/2}$, these two orientations have the same energy at low temperature.
However, since $\bar\epsilon_m$ still depends on $m$, the interface is not isotropic 
when  $\zeta=2^{-1/2}$ (actually when $m$ increases $\bar\epsilon_m$ increases and 
reaches a maximum value  $\bar\epsilon_2=\bar\epsilon_3=(2^{1/2}+2/3)5^{-1/5}$ for $m=2,3$, and then
decreases).  
This simple picture at zero temperature suggests that the most isotropic
line tension is obtained for $\zeta=2^{-1/2}$. However, we also see that 
the line-tension is $\bar\epsilon_m\approx (1/2+\zeta)+|\theta|/2$ for $\theta\rightarrow 0$.
Hence, the stiffness, which involves second order derivatives with respect to $\theta$,
diverges for $\theta=0$. As a consequence, the interface is always strongly anisotropic
at zero temperature around the (10) orientation in terms of stiffness.

\subsection{Finite temperature calculation}

\begin{figure}
\centering
\caption{Normalized line tension derived from the 2D Ising model. The normalized lattice constant is $\bar{a}=1$, and the normalized nearest-neighbor bond energies are $\bar{J}_x = \bar{J}_y = 1$.
The different color map (yellow is $\bar{\gamma}\approx 0$, darker for negative values) highlights where the line tension is negative, i.e.~for temperatures above the roughening temperature, $T_c$. 
The red line on the contour plot corresponds to $\bar{\gamma}\approx 0$.
Top panel: $\zeta=0$, $\overline{k_BT_c}(\theta=0)\approx 0.567$. 
Bottom panel: $\zeta = 1$, $\overline{k_BT_c}(\theta=0)\approx 1.345$; The inset c) shows small $T$ behavior ($\overline{k_BT}\leq 0.65$) for $\zeta= 1.5$. 
\label{fig:analitic_LT}}
\begin{subfigure}{\linewidth}
\includegraphics[width=\linewidth]{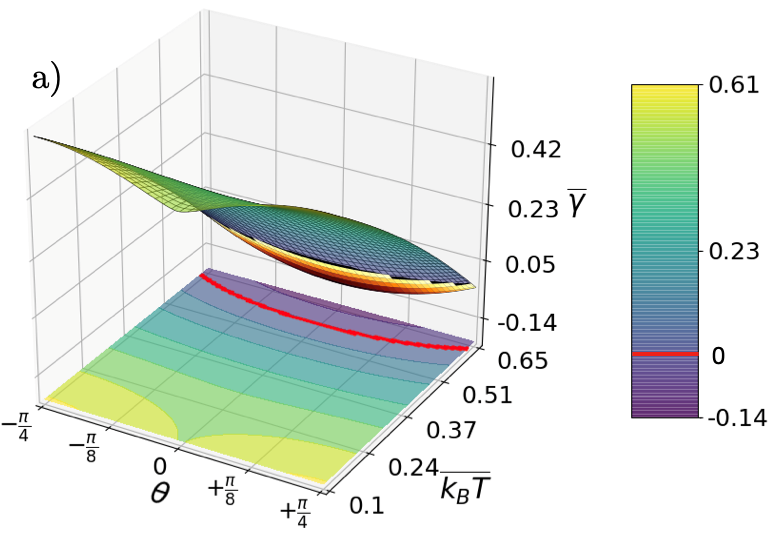}
\end{subfigure}
\begin{subfigure}{\linewidth}
\includegraphics[width=\linewidth]{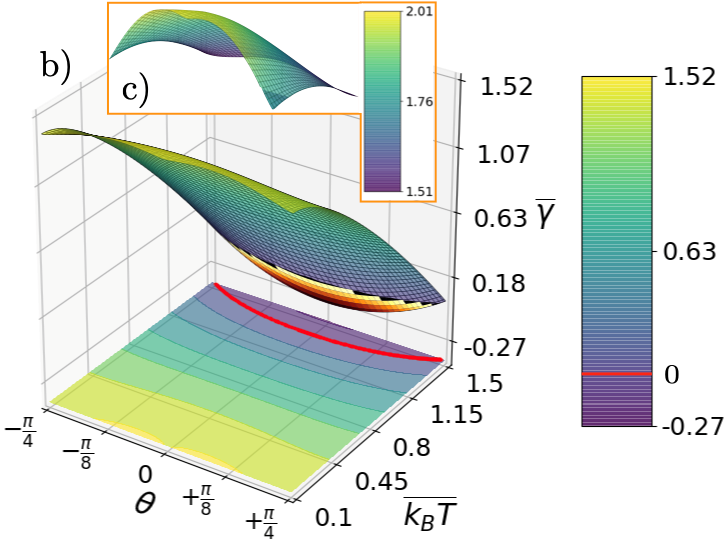}
\end{subfigure}
\end{figure}

We account for the orientation  $\theta$ by including a tilting field $H_t$. This is a standard approach which is discussed for instance in Ref.~\cite{Saito1996}.
Calculations similar to ours have been reported in the literature but only in the absence of next-nearest-neighbors or without accounting for arbitrary orientations \cite{Saito1996,Zandvliet2006,Ignacio2014,Zandvliet2015}.
However,  a relation for the interface free energy with next-nearest-neighbors was first derived in Ref.~\cite{Einstein2004}
within the same SOS approximation but employing a different procedure based on a constrained partition function.
Our result is equivalent to this previous result in the limit where the nearest-neighbor bonds in all directions are identical.
We briefly present the derivations in the main text. 
Further details are given in \cref{appendix:Ising}.

Instead of a single value $J_1$, we consider two distinct first-nearest-neighbor bond energies 
$J_x$ and $J_y$  along the $x$ and $y$ direction, respectively. 
The case $J_x\neq J_y$ allows one in principle to generalize the description to 2-fold anisotropy. This type of symmetry is relevant to some systems of technological importance such as Si(100) \cite{Ramstad1995}.
Once again, we use the notation  $a$ for the lattice constant. 
We use a coordinate system where $x$ is along the (10) direction,
while $y$ is along (01).
The interface is represented as a function $y(i)$ along $y$, at position $x = i\times a$, $i=1,\dots, N$.
With this definition, the interface exhibits no overhang along $x$ (solid-on-solid approximation).
Using these notations, the Hamiltonian of the system is
\begin{subequations}
\begin{align}
\label{eq:Hamiltonian}
\mathcal{H} =& \mathcal{H}_{nn} + \mathcal{H}_{nn'} + \mathcal{T}\, , \\
\mathcal{H}_{nn} =& \frac{J_y}{2} N + \frac{J_x}{2}\sum_{i=1}^{N}\frac{|y(i)-y(i-1)|}{a}\, , \\
\begin{split}
\mathcal{H}_{nn'} = &J_2 N + J_2 \sum_{i=1}^{N} (\frac{|y(i)-y(i-1)|}{a}-1)\\
&\hspace{1.5 cm} \times (1-\delta_{y(i)-y(i-1)})\, , 
\end{split}\\
\mathcal{T} =& - H_t\frac{y(N) - y(0)}{a}\, ,
\end{align}
\end{subequations}
where the different terms appearing in \cref{eq:Hamiltonian} are 
the contribution of first-nearest-neighbor interactions $\mathcal{H}_{nn}$, 
the contribution of second-nearest-neighbor interactions $\mathcal{H}_{nn'}$
and the contribution of the tilting field $\mathcal{T}$, 
which selects the average interface slope.

Defining the height difference between neighboring sites $n_i = y(i)-y(i-1)$, the energy per site reads
\begin{equation}
\label{eq:E_Ising}
E(n_i) =   J_y/2 + |n_i| J_x/2 + (|n_i| +\delta_{n_i}) J_2 - H_t n_i \, ,
\end{equation}
where we have used that $y(N) - y(0) = a\sum_{i=1}^{N} \left(y(i) - y(i-1)\right) $.
Note that the above is equivalent to the formation energy of a boundary of a 2D Ising lattice. In the following, we consider $J_2>0$, corresponding to ferromagnetic interactions \cite{Zandvliet2006}.

The partition function of the system is given by
\begin{equation}
\mathcal{Z} = \sum_{\{ n_i\}}e^{-E(n_i)/(k_BT)}\, .
\end{equation}
Since all sites are independent, we can drop the index $i$ and write $\mathcal{Z}$ in term of the partition function per site $z_0$
\begin{equation}
\label{eq:partitionFunc}
\mathcal{Z} = z_0^N\, ,
\end{equation}
with 
\begin{equation}
\label{eq:z0_implicit}
z_0 =\sum_{n=-\infty}^\infty e^{-E(n)/(k_BT)}\, .
\end{equation}
Let us introduce the equilibrium slope of the interface $p=\tan\theta$ (with respect to the (10) reference direction) defined as
\begin{align}
p & = \frac{1}{L}\langle\sum_{i=1}^{N} a n_i\rangle_{eq}= \frac{N}{L} a\langle n \rangle_{eq} = \langle n \rangle_{eq} 
\nonumber \\
&= \frac{1}{z_0}\sum_{n=-\infty}^\infty n e^{-E(n)/(k_BT)} \, ,
\end{align}
where $L=Na$ is the interface length along $x$.
Observing that $n$ is proportional to the tilting field $H_t$ in \cref{eq:E_Ising}, the above expression is equivalent to
\begin{equation}
\label{eq:p_implicit}
p=k_BT\partial_{H_t}\ln z_0\,.
\end{equation}
 Let us now consider the step or interface free energy per unit length along $x$ which, 
 assuming the other parameters fixed, is a function of the tilting field: 
\begin{align}
\hat{f}(H_t) = -\frac{k_BT}{L}\ln \mathcal{Z}=-\frac{k_BT}{a}\ln z_0\, .
\label{eq:def_fhat}
\end{align}
Combining   \cref{eq:p_implicit,eq:def_fhat}, we obtain
\begin{align}
\hat{f}'(H_t) = -\frac{p}{a}\,.
\label{eq:dfdHt_p}
\end{align}
It is convenient to rewrite the interface free energy
as a function of $p$ using the Legendre transformation \cite{Saito1996}:
\begin{equation}
\label{eq:Legendre}
f(p) = \hat{f}(H_t) - H_t\hat{f}'(H_t) = \hat{f}(H_t) + \frac{p}{a}H_t\, .
\end{equation}

From \cref{eq:dfdHt_p,eq:Legendre} we obtain an expression 
of the free energy density:
\begin{equation}
\label{eq:freeEn_Ht}
\begin{split}
\hat{f}(H_t) &= \frac{1}{a}(\frac{J_y}{2} + J_2) - \\
& \frac{k_BT}{a}\ln \left[1+ e^{-J_x/(2k_BT)}\frac{2\cosh(H_t/k_BT) - 2\alpha}{1-2\alpha\cosh(H_t/k_BT) + \alpha^2}\right]\, ,
\end{split}
\end{equation}
where $p$ and $H_t$ are related via
\begin{equation}
\label{eq:p}
\begin{split}
p = &\frac{2\sinh(H_t/k_BT)(1-\alpha^2)}{1-2\alpha\cosh(H_t/k_BT) + \alpha^2} \times\\
&\Bigl[ e^{J_x/(2k_BT)} \bigl(1 - 2 \alpha \cosh (H_t/k_BT)+\alpha^2\bigr)\\
& + 2\cosh(H_t/k_BT) -2\alpha \Bigr ]^{-1}\, ,
\end{split}
\end{equation}
with $\alpha = \exp [-(J_x/2 + J_2)/k_BT]$.
The detailed derivation of these equations is reported in \cref{appendix:Ising}.
Finally, as discussed in \cref{appendix:LT}, the angle dependent line tension and stiffness can be found from the following relations:
\begin{subequations}
	\label{eq:LT_all}
\begin{align}
\label{eq:LT}
\gamma(\theta) &= f(p)\cos\theta\, , \\
\label{eq:stiff}
\tilde{\gamma}(\theta) &=\frac{ f''(p)}{\cos^3\theta} = \frac{\partial_p H_t}{a\cos^3\theta}\, ,
\end{align}
\end{subequations}
where $\theta = \arctan(p)$.
The last equality  of \cref{eq:stiff} was obtained 
from the derivative of \cref{eq:Legendre} and using \cref{eq:dfdHt_p}.
In the limit where $J_x=J_y=J_1$, \cref{eq:LT_all,eq:freeEn_Ht,eq:p} are equivalent to the expressions found in Ref.~\cite{Einstein2004} \footnote{When comparing to the literature, care should be taken for possible integer factors in the bond energies, which are also sometimes indicated by $\epsilon$. Our notation is mainly inspired by Yukio Saito's book \cite{Saito1996}.}.

The relation \cref{eq:p} linking $p$ to $H_t$ cannot be inverted analytically.\footnote{ It is possible to derive explicit analytic expressions for the line tension and stiffness, under some approximations \cite{Einstein2007}. However, these are valid only up to $T\approx T_c/5$, where $T_c$ is the roughening temperature.}   
We thus resort to a numerical root finding method to obtain $H_t$ at a given slope $p$ and then compute $f(p)$ from \cref{eq:Legendre}
to determine $\gamma(\theta)$.
We also compute numerically $\partial_p H_t$ using finite differences
to obtain the stiffness $\tilde{\gamma}(\theta)$.

\begin{figure}
	\centering
	\caption{
		Normalized stiffness derived from the 2D Ising model.  The normalized lattice constant is $\bar{a}=1$, and the normalized nearest-neighbor bond energies are $\bar{J}_x = \bar{J}_y = 1$. Top panel: $\zeta=0$.
		Bottom panel: $\zeta = 1$. 
		The color map is in log scale.
		\label{fig:analitic_stiff}}
	\begin{subfigure}{\linewidth}
		\includegraphics[width=\linewidth]{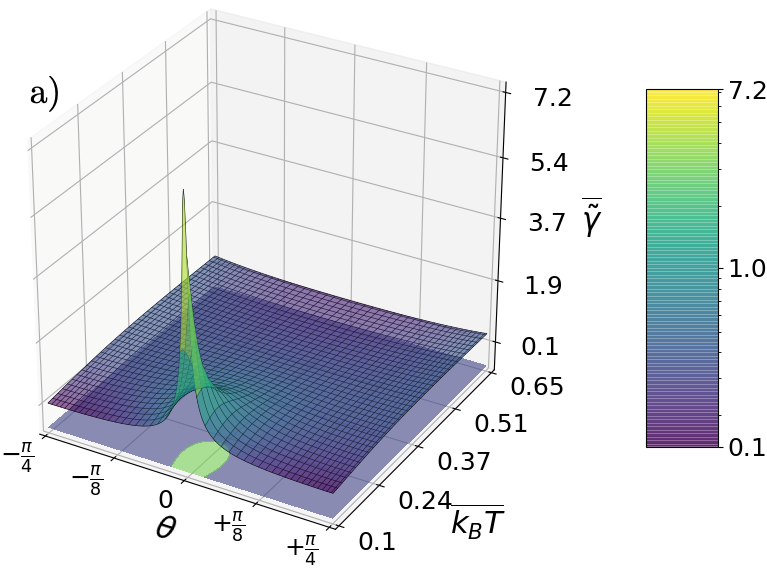}
	\end{subfigure}
	\begin{subfigure}{\linewidth}
		\includegraphics[width=\linewidth]{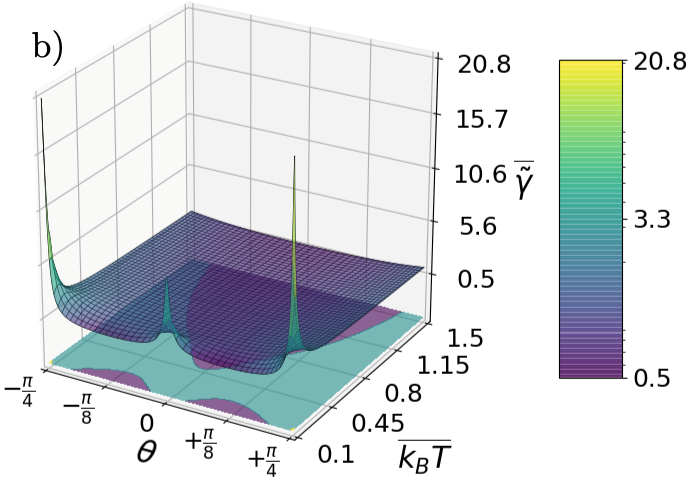}
	\end{subfigure}
\end{figure}

The results for the normalized line tension and stiffness are plotted in \cref{fig:analitic_LT,fig:analitic_stiff} with $J_x=J_y=J_1$ and $J_2 = \zeta J_1$. 
Let us focus first on the limit of sole first-nearest-neighbor interactions, $\zeta = 0$, represented in the top panels of \cref{fig:analitic_LT,fig:analitic_stiff}. 
For the line tension we recover the known limit of the classical 2D first-nearest-neighbor Ising model 
and equilibrium Wulff construction \cite{Rottman1981}: the zero temperature limit 
presents a characteristic cusp in correspondence of the (10) facet which  
flattens as the temperature increases.

As temperature increases we approach the roughening transition 
where the interface free energy vanishes. 
For $\zeta = 0$ (nearest-neighbor interactions only) and 
$\theta=0$ (corresponding to (10) orientation)  
we recover the standard result $k_BT_c = J_1 /[2\ln (1+\sqrt{2})] \approx0.567 J_1$ \cite{Saito1996,Einstein2007}. 
 
More generally, the roughening (critical) temperature at different orientations corresponds to the points where $\gamma(\theta) = 0$ in \cref{fig:analitic_LT}. 
In our model, the roughening temperature slightly increases as $\theta$ increases. 
This variation is unphysical, and results from the approximation
of an interface without overhangs along $x$. This description is not valid
close to the roughening temperature where statistical fluctuations are very large.
Our results at low temperatures (i.e.~far from the roughening transition) and for $\zeta = 0$ are 
also in qualitative agreement with the mean field approach reported in Ref.~\cite{Plapp1997}.
However, quantitative comparison is difficult given the scaling used in \cite{Plapp1997} which is based on a critical temperature relying on the mean-field approximation adopted.

When second-nearest-neighbor bonds do not vanish ($\zeta\neq 0$), 
the critical temperature $T_c$ increases linearly with $\zeta$,
as reported previously in  \cite{Zandvliet2006}.

The results for $\zeta\neq 0 $ are reported in the bottom panels of \cref{fig:analitic_LT,fig:analitic_stiff}.
As expected, when the temperature vanishes, 
the stiffness exhibits divergence  both at (10) and (11) orientations
\cite{Saito1996,Misbah2010,Rottman1981,Einstein2007,Einstein2015}. 
In contrast, as the temperature increases, the stiffness anisotropy decreases.

\section{Comparison to simulations: equilibrium fluctuations and stiffness}

\begin{table*}[t]
\centering
\caption{
Stiffness as obtained by KMC simulations, with $\bar{J}_1 =1$, $\overline{k_BT}=0.5$, and $\bar{L}=200$, compared to the analytic results obtained by solving and combining \cref{eq:freeEn_Ht,eq:p,eq:stiff} with the same parameters used in simulations. The simulations profile were convoluted by a Gaussian kernel with normalized standard-deviation
$\bar{\sigma}$. Results are averaged over $\bar{\sigma} \in[ 4,6,8]$.
\label{tab:summary}}
\resizebox{\textwidth}{!}{%
\begin{tabular}{@{}l|llll@{}}
\toprule
\diagbox{Orientation}{Bond ratio} & $\zeta = 0.7$ simulations & $\zeta = 0.7$ analytic & $\zeta = 1.4$ simulations & $\zeta = 1.4$ analytic \\ \midrule
(10) & $0.184\pm 0.003$ & $0.170$ & $0.238\pm 0.009$ & $0.218$ \\
(11) & $0.256\pm0.013$ & $0.242$ & $0.475\pm0.028$ & $0.532$ \\ \bottomrule
\end{tabular}%
}
\end{table*}

We compare the analytic results to KMC simulations focusing on 
a quantitative analysis of the stiffness. 	
The stiffness is related to the equilibrium roughness, a quantity that can be extracted from simulations.

The coordinate along the average direction of the step is denoted as $\xi$.
The interface position $h(\xi)$ is measured along the direction orthogonal to the $\xi$ direction.
The roughness $W$ is defined as the standard deviation of the interface position
\begin{equation}
\label{eq:roughness_cont}
W^2 (t) =\frac{1}{l}\int \mathrm{d}\xi\, h^2(\xi,t) - \frac{1}{l^2}\left (\int\mathrm{d}\xi \,h(\xi,t)\right) ^2\, ,
\end{equation}
with $l$ the length of the interface along $\xi$. 

In the continuum limit, the equilibrium roughness is related to the stiffness via
the well known formula \cite{Saito1994,Uwaha1992,Saito1996,Misbah2010}
\begin{equation}
\label{eq:rough_stiff}
\langle W^2\rangle_{eq} = \frac{ l k_BT}{12\tilde{\gamma}}\, .
\end{equation}
The above result is derived for periodic boundary conditions. 
A continuum  profile is obtained from our discrete lattice simulations as follows. First,
we define the solid phase characteristic function to be equal to $1$ in a square of size $a\times a$ 
around each lattice site occupied by a solid atom, and $0$ elsewhere.
We then perform a 2D Gaussian convolution
of the solid phase characteristic function with a 
length scale $\sigma$. Finally, we define the continuum interface profile as the line of height $1/2$
from the convoluted characteristic function. 
The length scale $\sigma$ therefore plays the same role as the resolution
of an experimental apparatus. This resolution ranges from
sub-atomic lengthcales in Scanning Tunneling Microscopy
to tens of nanometers in Low Energy Electron Microscopy,
and up to hundreds of nanometers in optical techniques \cite{Misbah2010}.

Since the roughness is dominated by long-wavelength modes, we expect the expression 
of the roughness, \cref{eq:rough_stiff}, to be 
accurate in the limit where we have a good scale separation 
$\sigma \ll l$. 
Nevertheless, $\sigma$ is finite in simulations or experiments
and its influence on our estimation of the roughness must be accounted for.
In order to do so, let us consider the convolution of the continuum profile along the $x$ direction (discarding the transversal direction):
\begin{equation}
	\bar{h}^\sigma(x,t) = \int \mathrm{d}x' \frac{1}{\sqrt{2\pi\sigma^2}}e^{-(x-x')^2/(2\sigma^2)} h(x',t)\, .
\end{equation}
In Fourier space, we have
\begin{equation}
	\bar{h}^\sigma_q = e^{-q^2\sigma^2/2}h_q \, ,
\end{equation}
with $q = 2\pi x/l$.
The static spectrum of the convoluted conitnuum profile is therefore
\begin{equation}
	\langle |\bar{h}^\sigma_q|^2  \rangle_{eq} =e^{-q^2\sigma^2} \langle |h_q|^2  \rangle_{eq}\, .
\end{equation}
From the equipartition of energy , one finds \cite{Saito1996,Misbah2010}
\begin{equation}
\langle |h_q|^2 \rangle_{eq} =\frac{k_BT}{\tilde{\gamma}q^2} l\, .
\end{equation}
The roughness of the convoluted profile is therefore calculated as
\begin{equation}
	\label{eq:roughness_fourier_corr}
	\begin{split}
		\langle W_\sigma^2 \rangle_{eq} &= \langle\frac{1}{l^2}\sum_n |h_q|^2 - \frac{1}{l^2}|h_{q=0}|^2\rangle_{eq}\\
		 &= l \frac{k_BT}{4\pi^2 \tilde{\gamma}}\, 2\sum_{n=1}^\infty\frac{e^{-4\pi^2\sigma^2 n^2/l^2}}{n^2} \,. \\
	\end{split}
\end{equation}
Defining
\begin{equation}
	\label{eq:S}
	S(\sigma/l) = \frac{6}{\pi ^2}\sum_{n=1}^\infty\frac{e^{-4\pi^2\sigma^2 n^2/l^2}}{n^2}\, ,
\end{equation}
we obtain
\begin{equation}
	\label{eq:rough_stiff_corr}
	\langle W_\sigma^2\rangle_{eq} = \frac{ l k_BT}{12\tilde{\gamma}} S(\sigma/l)\, .
\end{equation}
In the absence of convolution, $\sigma/l\rightarrow 0$ and $S(0) = 1$. We therefore recover
the previous expression of the roughness \cref{eq:rough_stiff}. 
When $\sigma/\ell \neq 0$, then $S(\sigma/\ell)<1$ and the roughness
of the convoluted continuum profile is decreased due to the elimination of 
the contribution of short wavelength modes. The convolution length scale
therefore plays the role of a short-wavelength cutoff.
As a consequence, the microscopic details of the models
that appear at the length scale $a$ will be irrelevant
for the convoluted roughness when $\sigma\gg a$. 
Hence, the convoluted continuum profile and the convoluted
simulation profile should be similar when  $\sigma\gg a$, 
and we should be able to use safely \cref{eq:rough_stiff_corr}
to any microscopic model when $\sigma\gg a$.
We therefore use \cref{eq:rough_stiff_corr} to extract the stiffness
from the measurement of the roughness of convoluted profiles.

In order to check this strategy, we have performed
KMC simulations at equilibrium for two arbitrary chosen values of the bond energy ratio, $\zeta = 0.7$ and $\zeta= 1.4$. We use a $200\times200$ square simulation box
with periodic boundary conditions comprising bands along  $(10)$ or $(11)$. 
We average the roughness over all interfaces in the simulation box,
and over time at equilibrium.
In \cref{app:roughness} we show simulation images and profile extractions together with the averaged roughness as a function of time (\cref{fig:bands,fig:roughness}). 

A direct extraction of the stiffness using \cref{eq:rough_stiff} on the 
convoluted profile provides an inconsistent value of $\tilde\gamma$ that depends on $\sigma$,
as shown in \cref{fig:deviation,fig:stiffSigma} in \cref{app:roughness}.
However, the roughness extracted from \cref{eq:rough_stiff_corr} is independent of $\sigma$
and is also in fair agreement with the predictions of the previous
section both for $(10)$ and $(11)$ interfaces.  
The use of  \cref{eq:rough_stiff_corr} then allows for a quantitative determination
of the stiffness as summarized in \cref{tab:summary}. 

We therefore propose that this strategy should be relevant
in experiments to determine the stiffness discarding the shortest
scales, that could exhibit short-wavelength deviations from the
continuum formula \cref{eq:rough_stiff}. These deviations could
either originate in system-dependent microscopic details of the physics of the interface, 
or on unavoidable distortions at the finest resolution scale in experimental images.

For illustrative purposes, in \cref{fig:instabilities}b  we show the effect of convolution, applied both on the solid island and the adatom cloud, on a general simulation frame. This allows to obtain a continuous density of adatoms which could be used for direct comparison with continuum models such as phase-field models \cite{Misbah2010,Karma1998}.



\section{Optimal bond ratio for isotropy}

\begin{figure}
	\centering
	\caption{Line tension anisotropy, measure by the relative standard deviation. Black dot: eye catcher showing the minimum at $\overline{k_BT}\approx 0.32$ and $\zeta\approx 0.54$. The color map is in log scale. \label{fig:minLT}}
	\includegraphics[width=\linewidth]{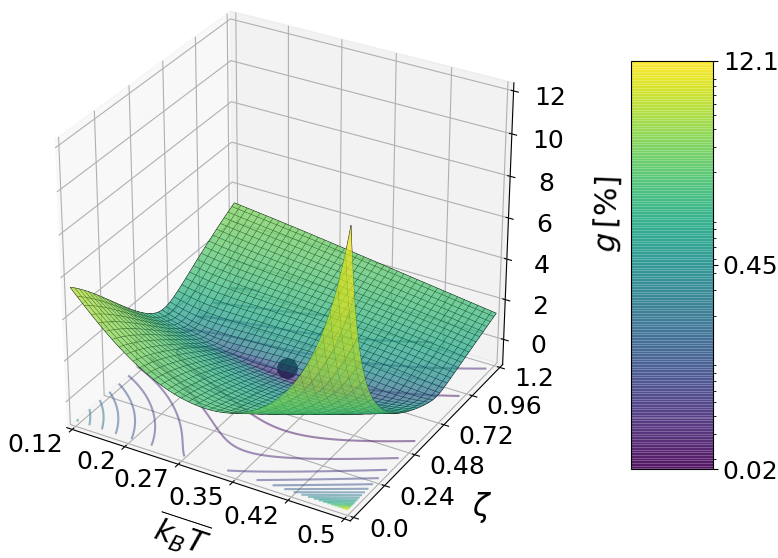}
\end{figure}

\begin{figure}
	\centering
	\caption{Stiffness anisotropy, measure by the relative standard deviation. Black dot: eye catcher showing the minimum at $\overline{k_BT}\approx 0.35$ and $\zeta\approx 0.43$. The color map is in log scale.
		\label{fig:minStiff}}
	\includegraphics[width=\linewidth]{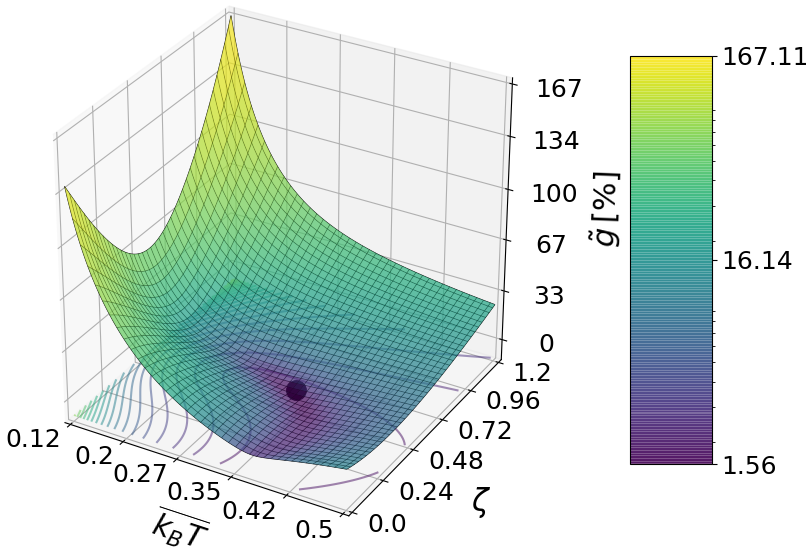}
\end{figure}

Let us introduce 
the averaged line tension\label{key} $\langle\gamma\rangle$ with respect to the orientation $\theta$:
\begin{equation}
\left\langle\gamma(T,\zeta)\right\rangle = 
\frac{4}{\pi}
\int_0^{\pi/4} \mathrm{d}\theta\, \gamma(\theta)\, .
\end{equation}
and its variance
\begin{equation}
\label{eq:variance}
v^2 (T,\zeta) = 
\frac{4}{\pi}
\int_0^{\pi/4} \mathrm{d}\theta\, \left (\gamma(\theta)-\langle\gamma\rangle\right )^2 ,
\end{equation}
The averages are taken up to $\theta = \pi/4$ due to symmetry.

We use the relative standard deviation 
of the line tension $g(T,\zeta) = v/\langle\gamma\rangle$ to
provide a measure of the anisotropy of the interface as a function of the temperature $T$ and bond energy ratio $\zeta$. 
Similarly, we define $\tilde{g}(T,\zeta)$ for the relative standard deviation of the stiffness.
In \cref{fig:minLT,fig:minStiff} we show 
$g$ and $\tilde g$ as a function of $\zeta$ and the normalized thermal energy $\overline{k_BT}$. 

In \cref{fig:minLT},  we observe that $g(\zeta_{\min})$ is of the order of $1\%$ in the range of (normalized) temperatures that we have explored.  
The highest level of isotropy is reached at $\overline{k_BT} \approx 0.32$ where $g \approx 0.016\%$ at $\zeta_{\min} \approx 0.54$.

The situation for the stiffness is different, with a more pronounced anisotropy.  
The most isotropic stiffness is reached again at a finite temperature 
$\overline{k_BT}\approx 0.35$ and $\zeta_{\min} \approx 0.43$ where $\tilde{g}\approx 1.56\%$.

\begin{figure}
\centering
\caption{Bond energy ratios realizing the highest isotropy levels ($\zeta_{\min}$) for line tension and stiffness, compared. The inset shows the correspondent relative standard deviation in log scale. The dip observed in the relative deviation associated to the stiffness at minimum (blue line) is around $2\%$ and corresponds to the violet area in \cref{fig:minStiff}. The relative standard deviation at minimum for the line tension is always below $1\%$. 
\label{fig:comparison}}
\includegraphics[width=\linewidth]{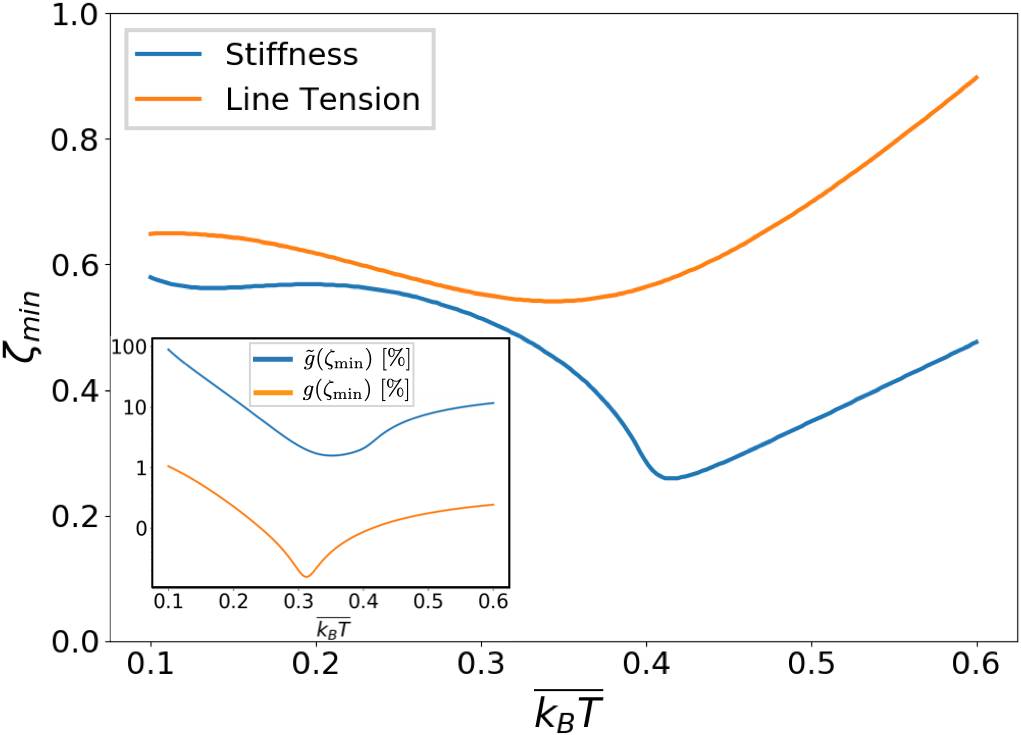}

\end{figure}

\Cref{fig:comparison} shows the value $\zeta_{\min}$ of the bond ratio 
that realizes the best degree of isotropy at a given temperature. The inset shows that the stiffness can
exhibit a small anisotropy only in a narrow temperature range between $\overline{k_BT}\approx 0.3$ and $\overline{k_BT}\approx 0.4$. The line tension, instead, can have small anisotropy in the whole temperature range explored (the relative variance at minimum is always below $1\%$). However, its highest isotropy is reached in a similar range of temperatures than the one observed for the stiffness.

Note that our model fails to describe the 
anisotropy at high temperatures close to the critical temperature.
As discussed above, the smallest critical temperature is reached
for $\zeta=0$ at $\overline{k_BT_c} (\zeta=0)\approx0.567$,
while the critical temperature increases significantly for larger $\zeta$ (see for instance \cref{fig:analitic_LT}b).
Close to the critical temperature, a spurious sharp increase
of the line tension anisotropy is seen in \cref{fig:minLT}.
This increase is due to the fact that the line tension vanishes
for some orientation before others in our model.
Since our observed minima of the  tension and stiffness anisotropy are far enough
from the transition, these minima might not be
a consequence of the spurious increase of the anisotropy 
at high temperature close to the transition. 
The accuracy of our results despite the SOS approximation 
is further supported by the good comparison to KMC simulations. 
Indeed, these simulations admit overhangs along the interface and are realized at a temperature 
closer to the transition (at $\overline{k_BT}=0.5$) 
than the minima (at $\overline{k_BT}\approx0.3$).

\section{Discussion}
\begin{figure*}[h]
	\centering
	
	\caption{
		Panel a): Initial condition (circular island).
		Panel b): Example of the convolution of a simulation frame. Left: unprocessed image, yellow: monolayer, red: adatoms. Right: Same image convoluted using smoothing parameter $\bar{\sigma}=2$ for the solid, and $\bar{\sigma}=6$ for the adatoms.
		Panels c) and d): Growth patterns obtained at $\overline{k_BT}=0.3$ with nearest-neighbor bond energy $\bar{J}_1=1$ in a regime of fast attachment (kinetic constant $\bar{A}=-3.5$).  
		The simulations are started with a circular seed (panel a)). The initial concentration of adatoms is $c = 0.15$ and the simulation box is $400\times 400$. Time flows from left to right.
		Top: Bond ratio realizing best degree of stiffness isotropy, $\zeta_{\min} (\overline{k_BT} = 0.3 ) = 0.51$.  
		Bottom: Higher bond ratio, $\zeta = 2$. 
		Other simulations parameters are: c) $\bar{E}_s = 1$; d) $\bar{E}_s = 3.5$; the images were convoluted with smoothing parameter $\bar{\sigma}=2$. 
		\label{fig:instabilities}}
	\begin{subfigure}{0.9\linewidth} 
		\includegraphics[width=\linewidth]{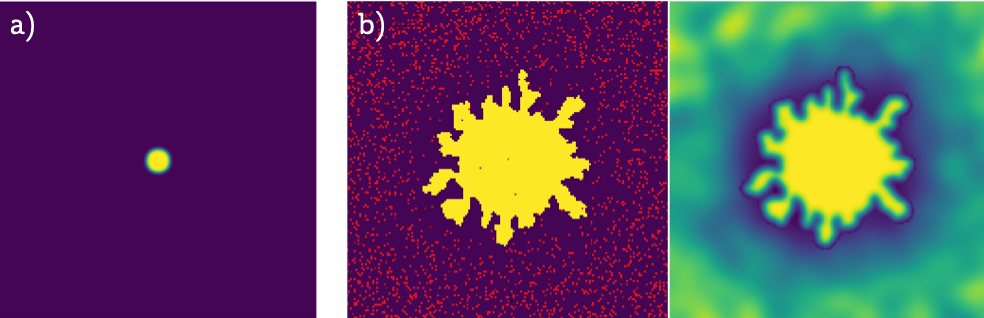}
	\end{subfigure}
	\vspace{2mm}
	
	\begin{subfigure}{0.9\linewidth}
		\includegraphics[width=\linewidth]{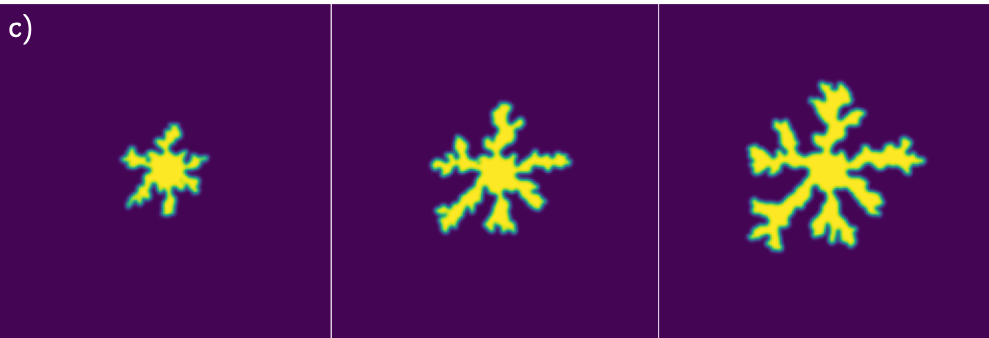}
	\end{subfigure}
	\vspace{2mm}
	
	\begin{subfigure}{0.9\linewidth}
		\includegraphics[width=\linewidth]{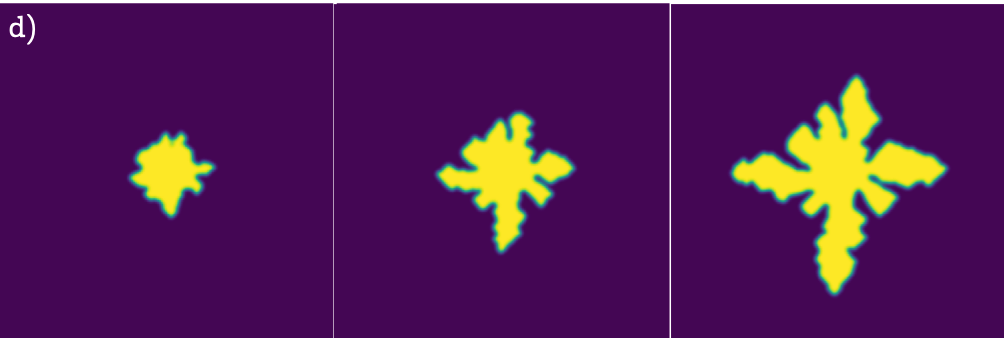}
	\end{subfigure}	
\end{figure*}

The role of crystal anisotropy in the selection of growth patterns has been extensively studied in the literature \cite{Ben-Jacob1990,Saito1987,Gouyet2003}. 
A detailed analysis of the effect of stiffness and line tension on the nonequilibrium patterns 
is beyond the scope of this work. However, it is interesting to show at least qualitatively the effect
of anisotropy on nonequilibrium patterns obtained within our KMC framework. 
In particular, in a diffusion driven regime ($\bar{A}<0$) we expect that isotropic stiffness should produce seaweed structures (tip splitting regime) 
whilst anisotropy could select specific dendritic patterns \cite{Ben-Jacob1990,Plapp1997,Saito1989}. 
While growth studies are usually performed with a constant supply of atoms
far from the crystal, we consider here the case where an initial concentration
larger from the equilibrium one is fixed in the initial conditions, leading to
a transient growth. On the one hand, since all atoms that contribute to the growth process must be present in the initial configuration, a large simulation box with many atoms is needed. These conditions are computationally more demanding than the constant growth scenario. Moreover, a high initial number of adatoms must be present in the system in order to observe instabilities. 
On the other hand, this growth mode could be well suited to describe some systems of technological relevance such as segregation of graphene on Ni substrate, which is based on temperature driven growth from a given concentration of carbon atoms \cite{Yu2008,Wu2014}.

We show in \cref{fig:instabilities}c and \cref{fig:instabilities}d growth patterns obtained by the KMC simulations at $\overline{k_BT}=0.3$. 
In the top panel we use the value $\zeta=\zeta_{\min} (\overline{k_BT}=0.3)=0.51$, which leads to the best degree of isotropy of the stiffness .
In the bottom panel a higher value, $\zeta = 2>\zeta_{\min}$, was taken.
As expected, the isotropic scenario produce seaweed patterns with the typical tip-splitting (see for instance fig.~2a in Ref.~\cite{Ben-Jacob1990}), while an anisotropic stiffness leads to anisotropic
dendrite-like patterns.
The dendrite tips are oriented in the (10) direction,
which corresponds to the direction of minimum stiffness \cite{Saito1994,Plapp1997}.
Videos of the full KMC simulations from which the frames used in \cref{fig:instabilities} were extracted are shown in the supplementary materials.

We chose to analyse 4-fold symmetry $J_x=J_y$ in this paper. In perspective, the KMC model here introduced is well suited to include the case $J_x\neq J_y$. This would allow one to address problems related to surface reconstruction where different rectangular-type lattice symmetries are adopted (e.g. Si(001) \cite{Ramstad1995}).
In addition, other type of lattice symmetries can be relevant for certain systems such as doped graphene \cite{Deretzis2014}.

\section{Conclusions}

Relating microscopic models to the macroscopic parameters of continuum models is a longstanding 
challenge for equilibrium and non-equilibrium interface dynamics. 
In this paper, we have established a quantitative relation between 2D monolayer islands 
with nearest and next-nearest-neighbor bonds, and the orientation dependent line tension and stiffness. This relation was
checked by means of kinetic Monte Carlo simulations.
In addition, we have proposed a method to extract the interface stiffness from simulations based on the
convolution of the interface profile by a Gaussian kernel. This method could also be useful
in the analysis of experimental data.
Furthermore, we provide a quantitative estimate of the degree of anisotropy of line tension and stiffness as a function of temperature and second-nearest-neighbor to nearest-neighbor bond strength. 
Reaching a good isotropy is an important condition in order
to test the predictions of continuum growth models, 
where anisotropy acts as a singular perturbation~\cite{Saito1996}.
Remarkably,  within a SOS description of a solid interface (no overhangs along the step), 
an optimum of isotropy is obtained at a finite temperature.
As the system approaches the roughening temperature, large statistical fluctuations hinder the validity of the model. 
Further theoretical and experimental evidence would be needed to support this non intuitive result.

We hope this work will contribute to partially fill the gap between the microscopic realm 
and continuous macroscopic theories and help to devise novel computer simulations approaches to the growth problem.

\clearpage

\appendix

\section*{Appendix}

\section{Line tension and stiffness}
\label{appendix:LT}

In this section we discuss some useful relations connecting the interface free energy to the line tension and stiffness. Note that the solid 2D monolayer interface (or in the Ising picture, the boundary between regions of opposite spins) is a one-dimensional function. The interface position is indicated by $h(x)$, where in the discrete limit $x= ia$ with $i$ an index running on the grid and $a$ the lattice constant.
The free energy of a step is defined as 
\begin{equation}
\label{eq:free_en0}
F[h] = \int \mathrm{ds}\,\, \gamma \left(\theta\right) = \int \sqrt{1+(\partial_x h)^2}\, \mathrm{dx}\,\, \gamma\left(\theta\right)\, ,
\end{equation}
with $s$ the arclength along the step (or in the Ising picture, along the phase boundary).
Expanding around a reference orientation, $\theta_0$ with $F[h,\theta=\theta_0]=F_0$, and considering the total free energy cost up to quadratic order we obtain
\begin{equation}
\label{eq:free_energy}
F[h] -F_0\approx \int \mathrm{dx}\, \frac{\tilde{\gamma}}{2}(\partial_x h) ^ 2 ,
\end{equation}
where we introduce the stiffness $\tilde{\gamma}(\theta) = \gamma(\theta) +\gamma''(\theta) $.

Let us now relate \cref{eq:free_en0} to  the slope $p = \tan \theta =\partial_x h$. Considering the free energy density per length $f(p) = F(p)/L$, for a uniform slope we have
\begin{equation}
Lf(p) = \int_0^L \mathrm{d}x \, \gamma(\theta)\sqrt{1+p^2} = L\gamma(\theta)\sqrt{1+p^2}\, ,
\end{equation}
where we have used that $\mathrm{d}s = \sqrt{1+(\mathrm{d}y/\mathrm{d})^2}\mathrm{d}x = \sqrt{1+p^2}\mathrm{d}x$.
It follows that the line tension is related to $f(p)$ by
\begin{equation}
\label{eq:LT_app}
\gamma(\theta) = \frac{f(p)}{\sqrt{1+p^2}} = f(p)\cos\theta \, .
\end{equation}
Using the definition of stiffness and that $\partial_\theta p = 1+p^2$,
by deriving the above relation we also obtain a general expression of the stiffness in terms of the free energy per unit length as a function of the slope :
\begin{equation}
\label{eq:stiff_app}
\tilde{\gamma}(\theta) =\frac{ f''(p)}{\cos^3\theta}\, .
\end{equation}

\section{2D Ising model}
\label{appendix:Ising}

Within the solid-on-solid (SOS) approximation, to our knowledge, the calculation of the interface free energy for arbitrary orientations with NN and NNN bonds was first carried out in Ref.~\cite{Einstein2004}.	
In the following, we show an alternative derivation. 
We start this section by introducing some useful relations and explicitly computing the partition function.
The main results of this section are \cref{eq:p,eq:freeEn_Ht} of the main text. 

As discussed in the main text the partition function is given by
\begin{equation}
\mathcal{Z} = z_0^ N \, ,
\end{equation}
where $N = L/a$ with $L$ the size of the step and $a$ the lattice constant, and
\begin{equation}
\label{eq:partition}
z_0 = \sum_{n=-\infty}^\infty e^{-E(n)/(k_BT)}\, ,
\end{equation}
the marginalized partition function.
The energy per site is given by
\begin{equation}
\label{eq:E_Ising_app}
\begin{split}
E(n) &= J_y/2 + J_2 + |n| J_x/2 + (|n| -1)(1-\delta_n)J_2 -H_t n \\
&=  J_y/2 + |n| J_x/2 + (|n| +\delta_n) J_2 - H_t n\, .
\end{split}
\end{equation}
From \cref{eq:partition} we can compute the average slope,
\begin{equation}
	\begin{split}
	\langle p\rangle_{eq} = &\frac{1}{L}\langle \sum_{i=1}^N a n_i\rangle_{eq}=\frac{a}{Na}\sum_{i=1}^{N} \langle n_i\rangle_{eq} = \frac{N}{N}\langle n\rangle_{eq}\\
	 &=  \frac{1}{\mathcal{Z}}\sum_{\{n_i\}} n_i \prod_{i=1}^{N}e^{-E(n_i)/(k_BT)}\, ,
	\end{split}
\end{equation}
where $i = 1,\dots , N$ and the sum is taken over all possible sites $i$ and values of $n_i$.
We can thus marginalize over all sites $k\neq i$ so that we obtain
\begin{equation}
\langle p\rangle_{eq} = \frac{1}{z_0}\sum_{n=-\infty}^\infty n e^{-E(n)/(k_BT)}\, .
\end{equation}
The above relation exploiting the proportionality between $n$ and $H_t$  in \cref{eq:E_Ising_app}, can be rewritten as
\begin{equation}
\label{eq:p_general}
\langle p\rangle_{eq}=k_BT\partial_{H_T}\ln z_0 \, .
\end{equation}
In the following, let us drop the ensemble average symbol on $p$.

As discussed in \cref{appendix:LT} we need an expression for the free energy per unit length as a function of the slope, $f(p)=F(p)/L$ .  
In principle, from the partition function $\mathcal{Z}$, we can compute $\hat{F} = -k_BT\ln \mathcal{Z}$. Observe that this will be a function of the Hamiltonian parameters and the temperature, $\hat{F} = \hat{F}(J_x,J_y,J_2,H_t,T)$. We hereby focus on the dependence to the tilting field, $H_t$. To move from $\hat{F}(H_t)$ to $F(p)$,
following \cite{Saito1996}, 
we start by noticing that $F(p)$ can be expressed by the Legendre transform
\begin{equation}
F(p) = \hat{F}(H_t) - H_t\partial_{H_t}\hat{F}(H_t)\, .
\end{equation}
Since $\hat{F}'(H_t) = -L/a\, p$, and using the definition of $f = F/L$, we have
\begin{equation}
\label{eq:legendre}
f(p) = \hat{f}(H_t)  +\frac{1}{a} p H_t\, .
\end{equation}
Finally, a last useful relation is obtained by deriving twice the above:
\begin{equation}
\label{eq:legendre_der}
f''(p) = \frac{1}{a} \partial_p H_t \, ,
\end{equation}
where again we used $\tilde{f}'(H_t) = -p/a$.
Admitting that we know the relation between $p$ and $H_t$, as discussed below, \cref{eq:legendre_der} together with \cref{eq:stiff_app} provide a way to obtain the stiffness at arbitrary orientations (\cref{eq:stiff} in the main text).

Let us now explicitly compute the marginalized partition function $z_0$ of \cref{eq:partition}.
Substituting $E(n)$ of \cref{eq:E_Ising_app}, we have
\begin{equation}
\begin{split}
z_0 &= e^{-J_y/(2k_BT)}\sum_{-\infty}^\infty e^{-\frac{1}{k_BT}\left[ J_x/2|n| + (|n| +\delta_n)J_2 -H_t n \right]}\\
&= e^{-J_y/(2k_BT)}\Bigl [ e^{-J_2/(k_BT)} - 2 \\
&+ \sum_{n=0}^\infty e^{-\frac{1}{k_BT}(J_x/2 + J_2 + H_T)n} + \sum_{n=0}^\infty e^{-\frac{1}{k_BT}(J_x/2 + J_2 - H_T)n}\Bigr ]\\
&= e^{-J_y/(2k_BT)}\Bigl [ e^{-J_2/k_BT} - 2 \\
&+ \frac{1}{1-e^{-(J_x/2 + J_2 + H_t)/(k_BT)}} + \frac{1}{1-e^{-(J_x/2 + J_2 - H_t)/(k_BT)}} \Bigr ]\,,
\end{split} 
\end{equation}
where we used the geometric series sum $\sum_{k=0}^\infty r^k = 1/(1-r)$ since $|r|<1$.
The above expression can be rewritten in a more compact way as 
\begin{equation}
\label{eq:z0_general}
\begin{split}
z_0 =& e^{-(J_y/2 + J_2)/(k_BT)} \bigl[ 1+ e^{-J_x/(2k_BT)}\\
&\times \frac{2\cosh(H_t/k_BT) - 2\alpha}{1-2\alpha\cosh(H_t/k_BT) +\alpha^2}\bigr ]\, ,
\end{split}
\end{equation}
where we introduced the term $\alpha = \exp [-(J_x/2 + J_2)/k_BT]$.
From the above equation we determine the free energy as a function of the tilting field (\cref{eq:freeEn_Ht} of the main text):
\begin{equation*}
\label{eq:freeEn_Ht_app}
\begin{split}
\hat{f}(H_t) &= \frac{1}{a}(\frac{J_y}{2} + J_2)- \frac{k_BT}{a}  \\
&\times\ln \left[1+ e^{-J_x/(2k_BT)}\frac{2\cosh(H_t/k_BT) - 2\alpha}{1-2\alpha\cosh(H_t/k_BT) + \alpha^2}\right].
\end{split}
\end{equation*}
Using \cref{eq:p_general}, we take the derivative of \cref{eq:z0_general} with respect to $H_t$, and obtain \cref{eq:p} of the main text:
\begin{equation*}
\begin{split}
p = &\frac{2\sinh(H_t/k_BT)(1-\alpha^2)}{1-2\alpha\cosh(H_t/k_BT) + \alpha^2}\\
& \times \Bigl[ e^{J_x/(2k_BT)} \bigl(1 - 2 \alpha \cosh (H_t/k_BT)+\alpha^2\bigr)\\
& + 2\cosh(H_t/k_BT) -2\alpha \Bigr ]^{-1}\, .
\end{split}
\end{equation*}
As discussed in the main text, this equation can be inverted using root finding methods.
In this work we employed the python \emph{SciPy} implementation of the Newton-Raphson method \cite{Scipy1.0}. 
Approximate explicit analytic solutions can be derived in the limit of $T\ll T_c$ \cite{Einstein2007,Einstein2004}.
Solutions are also available in limit scenarios, such as  $\theta=0$ and/or $J_2=0$ (only NN interactions) \cite{Rottman1981,Saito1996,Zandvliet2006,Ignacio2014,Zandvliet2015}. 
Finally, a noteworthy regime is that of small fluctuations limit
around the (10) orientation ($n\in [-1,0,1]$) \cite{Saito1996}, where the contribution of NNN vanishes.
\begin{figure}
	\centering
	\caption{Bias in the estimate of the stiffness (\cref{eq:rough_stiff_corr} in the main text) 
	as a function of the normalized smoothing parameter $\bar{\sigma}$. The simulation box is of size $\bar{L} =200$. Blue line: (10) step; Orange line: (11) step. \label{fig:deviation}}
	\includegraphics[width=\linewidth]{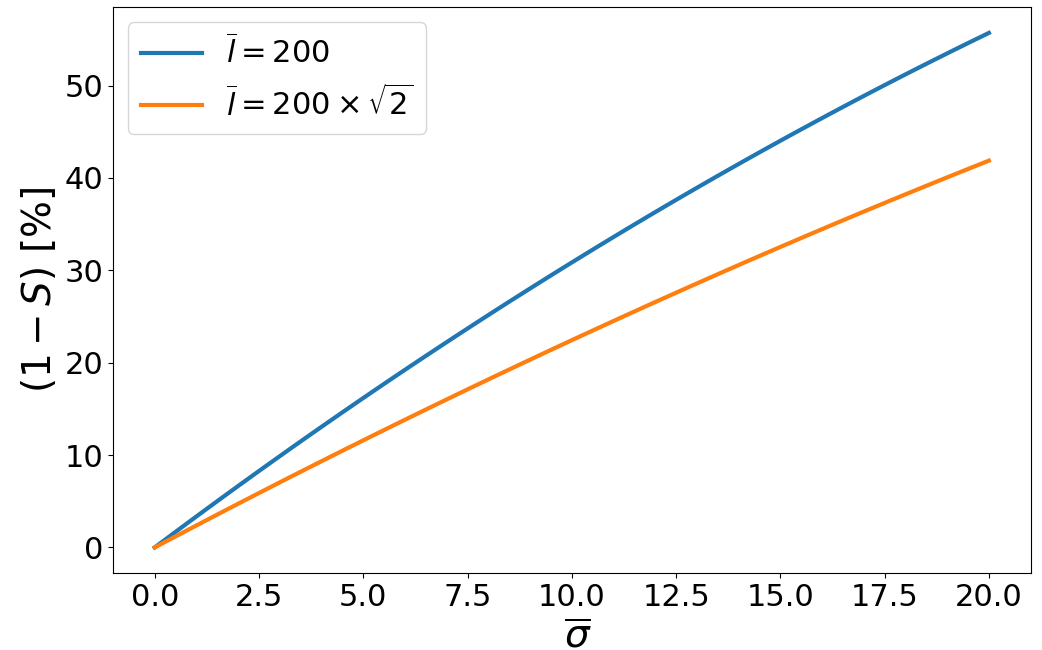}
\end{figure}
\begin{figure}[h!]
	\centering
	\caption{Stiffness as obtained by analytical calculations and KMC simulations at $\overline{k_BT} = 0.5$ for the directions $\theta=0$ (10) (top panel) and $\theta = \pi/4$ (11) (bottom panel), and for two values of the bond ratio $\zeta$, as a function of the normalized smoothing parameter $\bar{\sigma}$. We also show (red and yellow dots) results for \cref{eq:rough_stiff} not accounting for the correction term $S(\sigma/l)$ of \cref{eq:rough_stiff_corr}.
		The simulation box is $\bar{L}=200$. 
		\label{fig:stiffSigma}}
	\begin{subfigure}{\linewidth}
		\includegraphics[width =\linewidth]{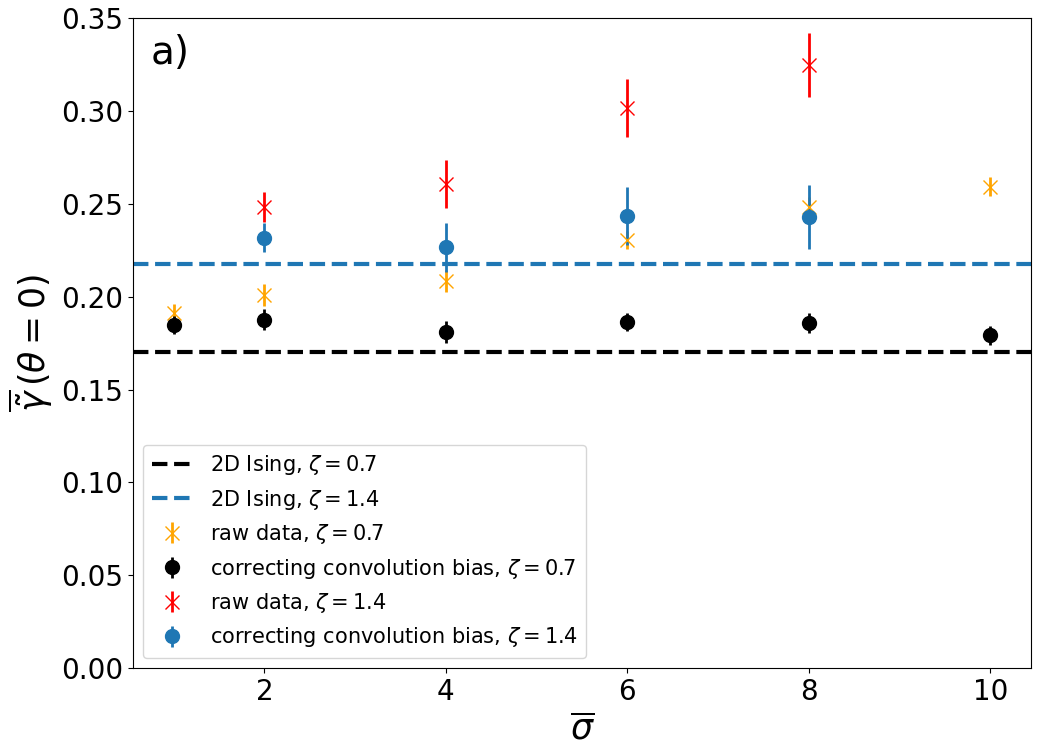}
	\end{subfigure}
	\begin{subfigure}{\linewidth}
		\includegraphics[width =\linewidth]{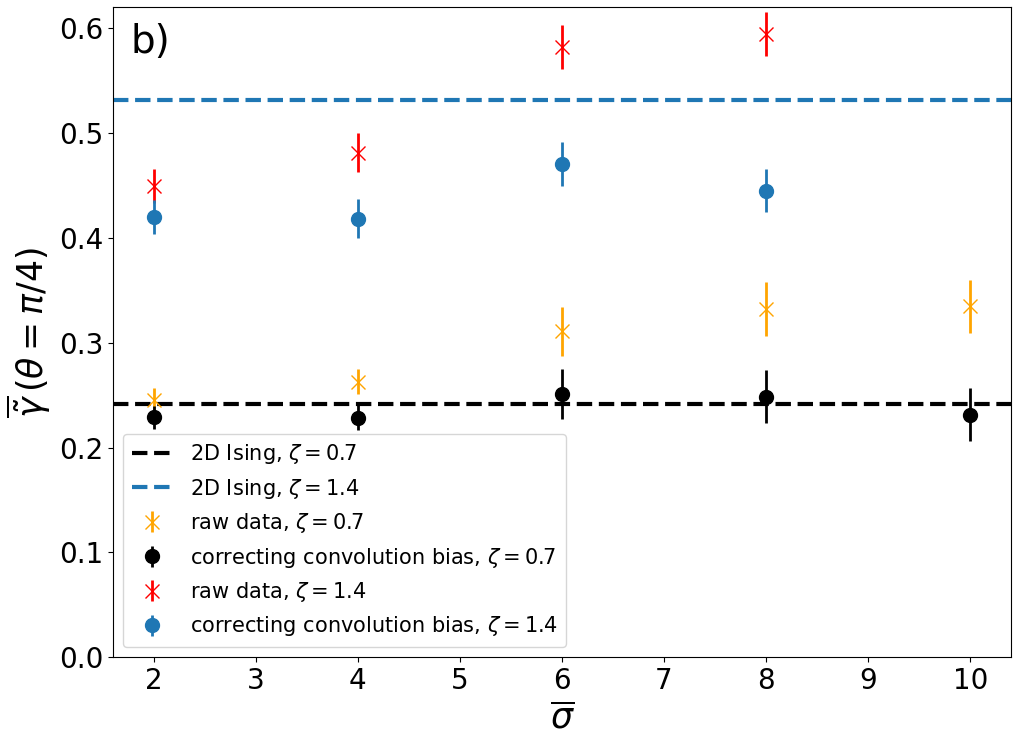}
	\end{subfigure}
\end{figure}


\subsection{Comparison to previous results}
To our knowledge, the first and only work providing a general expression for the orientation dependent free energy with second-nearest neighbors in the SOS approximation is Ref.~\cite{Einstein2004} (later re-discussed by the same authors in Ref.~\cite{Einstein2007}). In that paper the step free energy and equilibrium slope were obtained with a different derivation (using a constrained partition function and the saddle point approximation) but are completely equivalent to our expressions.
Indeed, the saddle point, $\rho_0$, plays the role of our (normalized) tilting field, $H_t/(k_BT)$. 
Recognizing that our free energy per unit length as a function of the tilting field $\hat{f}(H_t)$ corresponds to their $g(\rho_0)/a$, and using that (left hand side, Ref.~\cite{Einstein2004})
$H=\epsilon_k/(k_BT) = J_x/(2k_BT)$, $V=J_y/(2k_BT)$, $D=U=J_2/(2k_BT)$, $S=H+2D=J_x/(2k_BT) + J_2/(k_BT)$ so that $\alpha = \exp(-S)$, eqs.10 and A2 of Ref.~\cite{Einstein2004} correspond to \cref{eq:freeEn_Ht,eq:p} derived in this paper, respectively.


\section{Link between stiffness and roughness}
\label{app:roughness}
\begin{figure}
	\caption{ Normalized roughness of a (10) interface as a function of the normalized physical time, averaged over $n=12$ simulations for $\zeta = 0.7$, $\overline{k_BT}=0.5$, and using a smoothing parameter $\bar{\sigma} =6$. Each simulation contains three parallel bands. The shaded gray area represents the variance, the red line shows the time average taken from the asymptotic dynamics. \label{fig:roughness}} 
	\includegraphics[width = \linewidth]{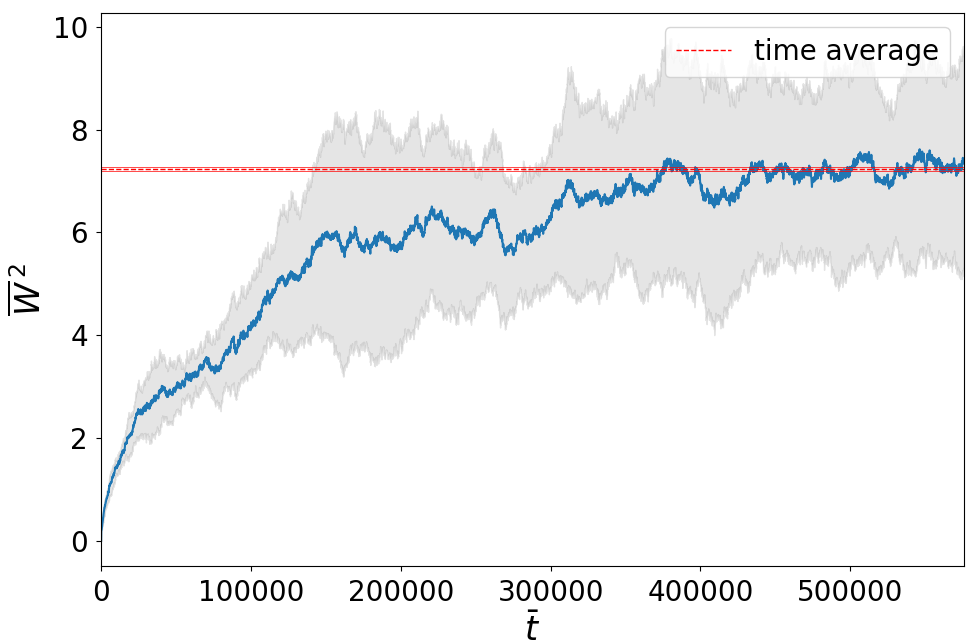}
\end{figure}
In this appendix, we compare the stiffness obtained from simulations to the analytic predictions.
Simulations parameters are: $\bar{L} =200$,  $\overline{k_BT}=0.1$, and $\bar{E}_s =1.5$.
In \cref{fig:deviation} we show the discrepancy,  $(1-S(\sigma/l))$, between the standard relation linking roughness and stiffness in \cref{eq:rough_stiff}, and \cref{eq:rough_stiff_corr} which accounts for the finite resolution of the observed interface. 
The term $S(\sigma/l)$ accounts for the decrease of the measured roughness due to the cut-off of short wavelength modes induced by the smoothing parameter (convolution length) $\sigma$.
We point out that this type of bias could be a source of error in experiments where the resolution is finite, if the correct relation, \cref{eq:rough_stiff_corr}, was not used. However, as observed in the figure, the bias also reduces as the step length increases.
In \cref{fig:stiffSigma} we show the stiffness as obtained by simulations compared to the analytic prediction from the 2D Ising lattice calculations. To further illustrate the importance of the correction $S(\sigma/l)$ to \cref{eq:rough_stiff}, we also display raw results without the application of $S(\sigma/l)$. Apart from a poor comparison to the analytic predictions (dashed lines), these raw results also depend on $\sigma$. 

\begin{figure}
	\centering
	\caption{Example of typical simulation images used to extract the roughness of the interface. These images are obtained after a Gaussian convolution at $\bar{\sigma} =4 $ and the simulation box is $200 \times 200$. The interface position (black line) is extracted from a homemade python script which finds the position of the $1/2$ level set.
	Top panel: (10) orientation.
	Bottom panel, b): (11) orientation; c): Interface profile extracted after rotating b) by $\theta = 45^o$.
		\label{fig:bands} }
	\begin{subfigure}{\linewidth}
		\includegraphics[width =\linewidth]{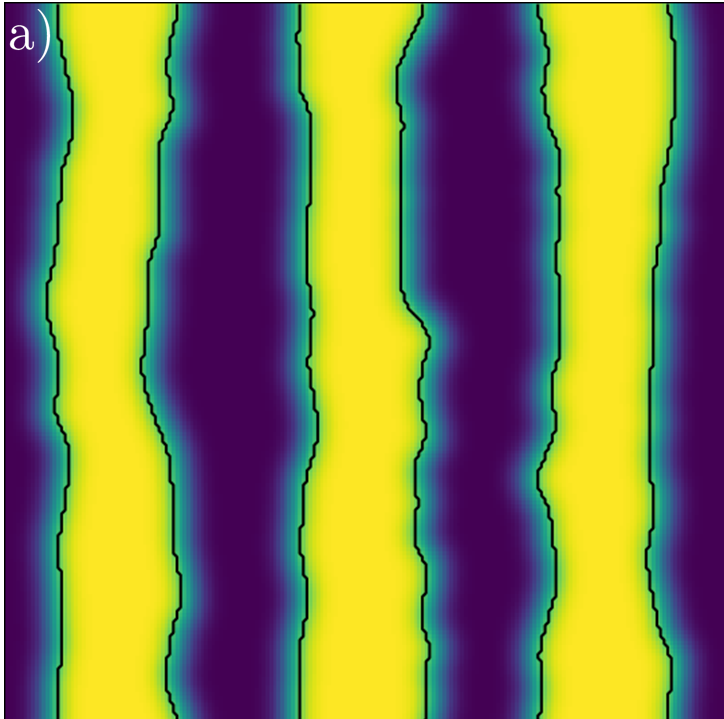}
	\end{subfigure}
	\begin{subfigure}{\linewidth}
		\includegraphics[width =\linewidth]{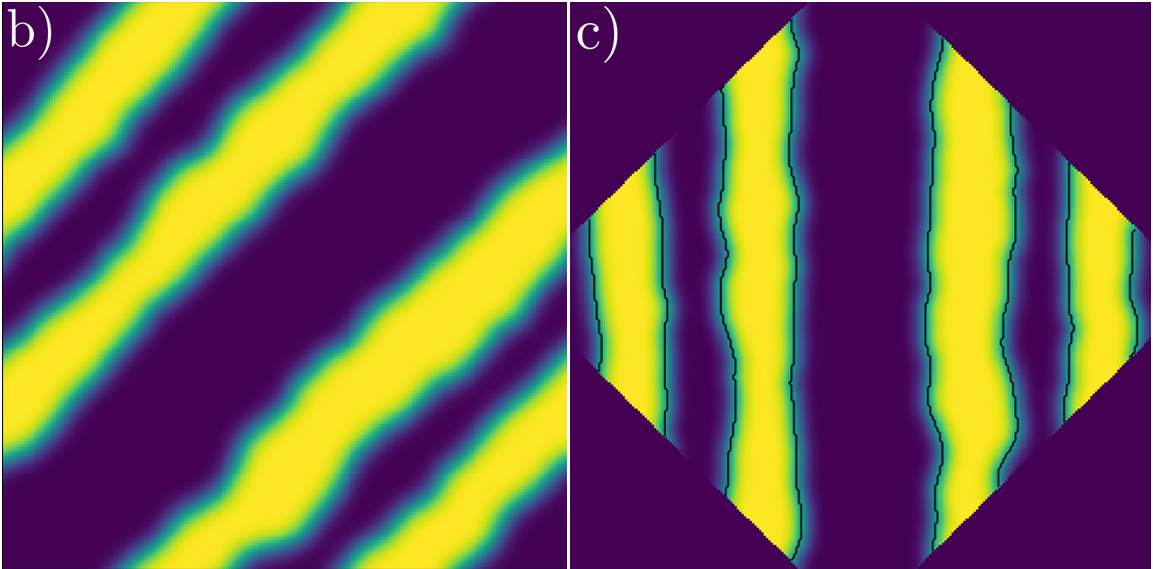}
	\end{subfigure}
	
\end{figure}

Finally in \cref{fig:bands,fig:roughness}, we illustrate the procedure followed to extract the roughness from KMC simulations. 
To measure the roughness along (11) we simulated periodic parallel bands rotated at $\theta = 45^o$. The profile is then extracted performing a $45^o$ rotation of the images and using the same routine employed on the (10) facet to extract the solid interface.

\clearpage

\bibliographystyle{cas-model2-names}

\RaggedRight 
\bibliography{kmc_biblio}

\begin{thebibliography}{31}
\expandafter\ifx\csname natexlab\endcsname\relax\def\natexlab#1{#1}\fi
\providecommand{\url}[1]{\texttt{#1}}
\providecommand{\href}[2]{#2}
\providecommand{\path}[1]{#1}
\providecommand{\DOIprefix}{doi:}
\providecommand{\ArXivprefix}{arXiv:}
\providecommand{\URLprefix}{URL: }
\providecommand{\Pubmedprefix}{pmid:}
\providecommand{\doi}[1]{\href{http://dx.doi.org/#1}{\path{#1}}}
\providecommand{\Pubmed}[1]{\href{pmid:#1}{\path{#1}}}
\providecommand{\bibinfo}[2]{#2}
\ifx\xfnm\relax \def\xfnm[#1]{\unskip,\space#1}\fi
\bibitem[{Barab{\'a}si and Stanley(1995)}]{Barabasi1995}
\bibinfo{author}{Barab{\'a}si, A.L.}, \bibinfo{author}{Stanley, H.E.},
  \bibinfo{year}{1995}.
\newblock \bibinfo{title}{Fractal Concepts in Surface Growth}.
\newblock \bibinfo{publisher}{{Cambridge University Press}},
  \bibinfo{address}{{New York, NY, USA}}.
\bibitem[{{Ben-Jacob} and Garik(1990)}]{Ben-Jacob1990}
\bibinfo{author}{{Ben-Jacob}, E.}, \bibinfo{author}{Garik, P.},
  \bibinfo{year}{1990}.
\newblock \bibinfo{title}{The formation of patterns in non-equilibrium growth}.
\newblock \bibinfo{journal}{Nature} \bibinfo{volume}{343},
  \bibinfo{pages}{523--530}.
\newblock \DOIprefix\doi{10.1038/343523a0}.
\bibitem[{Chame et~al.(2020)Chame, Saito and {Pierre-Louis}}]{Chame2020}
\bibinfo{author}{Chame, A.}, \bibinfo{author}{Saito, Y.},
  \bibinfo{author}{{Pierre-Louis}, O.}, \bibinfo{year}{2020}.
\newblock \bibinfo{title}{Orientation and morphology of solid-state dewetting
  holes}.
\newblock \bibinfo{journal}{Physical Review Materials} \bibinfo{volume}{4},
  \bibinfo{pages}{094006}.
\newblock \DOIprefix\doi{10.1103/PhysRevMaterials.4.094006}.
\bibitem[{Curiotto et~al.(2019)Curiotto, M{\"u}ller, {El-Barraj}, Cheynis,
  {Pierre-Louis} and Leroy}]{Curiotto2019}
\bibinfo{author}{Curiotto, S.}, \bibinfo{author}{M{\"u}ller, P.},
  \bibinfo{author}{{El-Barraj}, A.}, \bibinfo{author}{Cheynis, F.},
  \bibinfo{author}{{Pierre-Louis}, O.}, \bibinfo{author}{Leroy, F.},
  \bibinfo{year}{2019}.
\newblock \bibinfo{title}{{{2D}} nanostructure motion on anisotropic surfaces
  controlled by electromigration}.
\newblock \bibinfo{journal}{Applied Surface Science} \bibinfo{volume}{469},
  \bibinfo{pages}{463--470}.
\newblock \DOIprefix\doi{10.1016/j.apsusc.2018.11.049}.
\bibitem[{Danker et~al.(2004)Danker, {Pierre-Louis}, Kassner and
  Misbah}]{Danker2004}
\bibinfo{author}{Danker, G.}, \bibinfo{author}{{Pierre-Louis}, O.},
  \bibinfo{author}{Kassner, K.}, \bibinfo{author}{Misbah, C.},
  \bibinfo{year}{2004}.
\newblock \bibinfo{title}{Peculiar {{Effects}} of {{Anisotropic Diffusion}} on
  {{Dynamics}} of {{Vicinal Surfaces}}}.
\newblock \bibinfo{journal}{Physical Review Letters} \bibinfo{volume}{93},
  \bibinfo{pages}{185504}.
\newblock \DOIprefix\doi{10.1103/PhysRevLett.93.185504}.
\bibitem[{Deretzis and La~Magna(2014)}]{Deretzis2014}
\bibinfo{author}{Deretzis, I.}, \bibinfo{author}{La~Magna, A.},
  \bibinfo{year}{2014}.
\newblock \bibinfo{title}{Origin and impact of sublattice symmetry breaking in
  nitrogen-doped graphene}.
\newblock \bibinfo{journal}{Physical Review B} \bibinfo{volume}{89},
  \bibinfo{pages}{115408}.
\newblock \DOIprefix\doi{10.1103/PhysRevB.89.115408}.
\bibitem[{Einstein(2015)}]{Einstein2015}
\bibinfo{author}{Einstein, T.}, \bibinfo{year}{2015}.
\newblock \bibinfo{title}{Equilibrium {{Shape}} of {{Crystals}}}, in:
  \bibinfo{booktitle}{Handbook of {{Crystal Growth}}}.
  \bibinfo{publisher}{{Elsevier}}, pp. \bibinfo{pages}{215--264}.
\newblock \DOIprefix\doi{https://arxiv.org/abs/1501.02213}.
\bibitem[{Gouyet et~al.(2003)Gouyet, Plapp, Dieterich and Maass}]{Gouyet2003}
\bibinfo{author}{Gouyet, J.F.}, \bibinfo{author}{Plapp, M.},
  \bibinfo{author}{Dieterich, W.}, \bibinfo{author}{Maass, P.},
  \bibinfo{year}{2003}.
\newblock \bibinfo{title}{Description of far-from-equilibrium processes by
  mean-field lattice gas models}.
\newblock \bibinfo{journal}{Advances in Physics} \bibinfo{volume}{52},
  \bibinfo{pages}{523--638}.
\newblock \DOIprefix\doi{10.1080/00018730310001615932}.
\bibitem[{Ignacio(2014)}]{Ignacio2014}
\bibinfo{author}{Ignacio, M.}, \bibinfo{year}{2014}.
\newblock \bibinfo{title}{{\'Etude th\'eorique du mouillage de nano-cristaux
  solides sur des substrats nano-pattern\'es}}.
\newblock Ph.D. thesis. Universit\'e Claude Bernard Lyon 1.
  \bibinfo{address}{{Lyon - France}}.
\newblock \DOIprefix\doi{https://tel.archives-ouvertes.fr/tel-01127957}.
\bibitem[{Karma and Rappel(1998)}]{Karma1998}
\bibinfo{author}{Karma, A.}, \bibinfo{author}{Rappel, W.J.},
  \bibinfo{year}{1998}.
\newblock \bibinfo{title}{Quantitative phase-field modeling of dendritic growth
  in two and three dimensions}.
\newblock \bibinfo{journal}{Physical Review E} \bibinfo{volume}{57},
  \bibinfo{pages}{4323--4349}.
\newblock \DOIprefix\doi{10.1103/PhysRevE.57.4323}.
\bibitem[{Kotrla(1996)}]{Kotrla1996}
\bibinfo{author}{Kotrla, M.}, \bibinfo{year}{1996}.
\newblock \bibinfo{title}{Numerical simulations in the theory of crystal
  growth}.
\newblock \bibinfo{journal}{Computer Physics Communications}
  \bibinfo{volume}{97}, \bibinfo{pages}{82--100}.
\newblock \DOIprefix\doi{10.1016/0010-4655(96)00023-9}.
\bibitem[{Krishnamachari et~al.(1996)Krishnamachari, McLean, Cooper and
  Sethna}]{Krishnamachari1996}
\bibinfo{author}{Krishnamachari, B.}, \bibinfo{author}{McLean, J.},
  \bibinfo{author}{Cooper, B.}, \bibinfo{author}{Sethna, J.},
  \bibinfo{year}{1996}.
\newblock \bibinfo{title}{Gibbs-{{Thomson}} formula for small island sizes:
  Corrections for high vapor densities}.
\newblock \bibinfo{journal}{Physical Review B} \bibinfo{volume}{54},
  \bibinfo{pages}{8899--8907}.
\newblock \DOIprefix\doi{10.1103/PhysRevB.54.8899}.
\bibitem[{Maksym(1988)}]{Maksym1988}
\bibinfo{author}{Maksym, P.A.}, \bibinfo{year}{1988}.
\newblock \bibinfo{title}{Fast {{Monte Carlo}} simulation of {{MBE}} growth}.
\newblock \bibinfo{journal}{Semiconductor Science and Technology}
  \bibinfo{volume}{3}, \bibinfo{pages}{594--596}.
\newblock \DOIprefix\doi{10.1088/0268-1242/3/6/014}.
\bibitem[{Misbah et~al.(2010)Misbah, {Pierre-Louis} and Saito}]{Misbah2010}
\bibinfo{author}{Misbah, C.}, \bibinfo{author}{{Pierre-Louis}, O.},
  \bibinfo{author}{Saito, Y.}, \bibinfo{year}{2010}.
\newblock \bibinfo{title}{Crystal surfaces in and out of equilibrium: A modern
  view}.
\newblock \bibinfo{journal}{Reviews of Modern Physics} \bibinfo{volume}{82},
  \bibinfo{pages}{981--1040}.
\newblock \DOIprefix\doi{10.1103/RevModPhys.82.981}.
\bibitem[{{Pierre-Louis} and Einstein(2000)}]{Pierre-Louis2000}
\bibinfo{author}{{Pierre-Louis}, O.}, \bibinfo{author}{Einstein, T.L.},
  \bibinfo{year}{2000}.
\newblock \bibinfo{title}{Electromigration of single-layer clusters}.
\newblock \bibinfo{journal}{Physical Review B} \bibinfo{volume}{62},
  \bibinfo{pages}{10}.
\newblock \DOIprefix\doi{10.1103/PhysRevB.62.13697}.
\bibitem[{Plapp and Gouyet(1997)}]{Plapp1997}
\bibinfo{author}{Plapp, M.}, \bibinfo{author}{Gouyet, J.F.},
  \bibinfo{year}{1997}.
\newblock \bibinfo{title}{Dendritic growth in a mean-field lattice gas model}.
\newblock \bibinfo{journal}{Physical Review E} \bibinfo{volume}{55},
  \bibinfo{pages}{45--57}.
\newblock \DOIprefix\doi{10.1103/PhysRevE.55.45}.
\bibitem[{Ramstad et~al.(1995)Ramstad, Brocks and Kelly}]{Ramstad1995}
\bibinfo{author}{Ramstad, A.}, \bibinfo{author}{Brocks, G.},
  \bibinfo{author}{Kelly, P.J.}, \bibinfo{year}{1995}.
\newblock \bibinfo{title}{Theoretical study of the {{Si}}(100) surface
  reconstruction}.
\newblock \bibinfo{journal}{Physical Review B} \bibinfo{volume}{51},
  \bibinfo{pages}{14504--14523}.
\newblock \DOIprefix\doi{10.1103/PhysRevB.51.14504}.
\bibitem[{Rottman and Wortis(1981)}]{Rottman1981}
\bibinfo{author}{Rottman, C.}, \bibinfo{author}{Wortis, M.},
  \bibinfo{year}{1981}.
\newblock \bibinfo{title}{Exact equilibrium crystal shapes at nonzero
  temperature in two dimensions}.
\newblock \bibinfo{journal}{Physical Review B} \bibinfo{volume}{24},
  \bibinfo{pages}{6274--6277}.
\newblock \DOIprefix\doi{10.1103/PhysRevB.24.6274}.
\bibitem[{Saito(1996)}]{Saito1996}
\bibinfo{author}{Saito, Y.}, \bibinfo{year}{1996}.
\newblock \bibinfo{title}{Statistical Physics of Crystal Growth}.
\newblock \bibinfo{publisher}{{World Scientific}}, \bibinfo{address}{{River
  Edge, N.J}}.
\bibitem[{Saito et~al.(1987)Saito, {Goldbeck-Wood} and
  {M{\"u}ller-Krumbhaar}}]{Saito1987}
\bibinfo{author}{Saito, Y.}, \bibinfo{author}{{Goldbeck-Wood}, G.},
  \bibinfo{author}{{M{\"u}ller-Krumbhaar}, H.}, \bibinfo{year}{1987}.
\newblock \bibinfo{title}{Dendritic crystallization: Numerical study of the
  one-sided model}.
\newblock \bibinfo{journal}{Physical Review Letters} \bibinfo{volume}{58},
  \bibinfo{pages}{1541--1543}.
\newblock \DOIprefix\doi{10.1103/PhysRevLett.58.1541}.
\bibitem[{Saito and Ueta(1989)}]{Saito1989}
\bibinfo{author}{Saito, Y.}, \bibinfo{author}{Ueta, T.}, \bibinfo{year}{1989}.
\newblock \bibinfo{title}{Monte {{Carlo}} studies of equilibrium and growth
  shapes of a crystal}.
\newblock \bibinfo{journal}{Physical Review A} \bibinfo{volume}{40},
  \bibinfo{pages}{3408--3419}.
\newblock \DOIprefix\doi{10.1103/PhysRevA.40.3408}.
\bibitem[{Saito and Uwaha(1994)}]{Saito1994}
\bibinfo{author}{Saito, Y.}, \bibinfo{author}{Uwaha, M.}, \bibinfo{year}{1994}.
\newblock \bibinfo{title}{Fluctuation and instability of steps in a diffusion
  field}.
\newblock \bibinfo{journal}{Physical Review B} \bibinfo{volume}{49},
  \bibinfo{pages}{10677--10692}.
\newblock \DOIprefix\doi{10.1103/PhysRevB.49.10677}.
\bibitem[{Stasevich and Einstein(2007)}]{Einstein2007}
\bibinfo{author}{Stasevich, T.J.}, \bibinfo{author}{Einstein, T.L.},
  \bibinfo{year}{2007}.
\newblock \bibinfo{title}{Analytic {{Formulas}} for the {{Orientation
  Dependence}} of {{Step Stiffness}} and {{Line Tension}}: Key {{Ingredients}}
  for {{Numerical Modeling}}}.
\newblock \bibinfo{journal}{Multiscale Modeling \& Simulation}
  \bibinfo{volume}{6}, \bibinfo{pages}{90--104}.
\newblock \DOIprefix\doi{10.1137/060662861},
  \href{http://arxiv.org/abs/cond-mat/0609237}{\tt arXiv:cond-mat/0609237}.
\bibitem[{Stasevich et~al.(2004)Stasevich, Einstein, Zia, Giesen, Ibach and
  Szalma}]{Einstein2004}
\bibinfo{author}{Stasevich, T.J.}, \bibinfo{author}{Einstein, T.L.},
  \bibinfo{author}{Zia, R.K.P.}, \bibinfo{author}{Giesen, M.},
  \bibinfo{author}{Ibach, H.}, \bibinfo{author}{Szalma, F.},
  \bibinfo{year}{2004}.
\newblock \bibinfo{title}{Effects of next-nearest-neighbor interactions on the
  orientation dependence of step stiffness: Reconciling theory with experiment
  for {{Cu}}(001)}.
\newblock \bibinfo{journal}{Physical Review B} \bibinfo{volume}{70},
  \bibinfo{pages}{245404}.
\newblock \DOIprefix\doi{10.1103/PhysRevB.70.245404}.
\bibitem[{Steimer et~al.(2001)Steimer, Giesen, Verheij and Ibach}]{Steimer2001}
\bibinfo{author}{Steimer, C.}, \bibinfo{author}{Giesen, M.},
  \bibinfo{author}{Verheij, L.}, \bibinfo{author}{Ibach, H.},
  \bibinfo{year}{2001}.
\newblock \bibinfo{title}{Experimental determination of step energies from
  island shape fluctuations: A comparison to the equilibrium shape method for
  {{Cu}}(100), {{Cu}}(111), and {{Ag}}(111)}.
\newblock \bibinfo{journal}{Physical Review B} \bibinfo{volume}{64},
  \bibinfo{pages}{085416}.
\newblock \DOIprefix\doi{10.1103/PhysRevB.64.085416}.
\bibitem[{Uwaha and Saito(1992)}]{Uwaha1992}
\bibinfo{author}{Uwaha, M.}, \bibinfo{author}{Saito, Y.}, \bibinfo{year}{1992}.
\newblock \bibinfo{title}{Kinetic smoothing and roughening of a step with
  surface diffusion}.
\newblock \bibinfo{journal}{Physical Review Letters} \bibinfo{volume}{68},
  \bibinfo{pages}{224--227}.
\newblock \DOIprefix\doi{10.1103/PhysRevLett.68.224}.
\bibitem[{Virtanen and {et al.}(2020)}]{Scipy1.0}
\bibinfo{author}{Virtanen}, \bibinfo{author}{{et al.}}, \bibinfo{year}{2020}.
\newblock \bibinfo{title}{{{SciPy}} 1.0: Fundamental algorithms for scientific
  computing in {{Python}}}.
\newblock \bibinfo{journal}{Nature Methods} \bibinfo{volume}{17},
  \bibinfo{pages}{261--272}.
\newblock \DOIprefix\doi{10.1038/s41592-019-0686-2}.
\bibitem[{Wu et~al.(2014)Wu, Zhang, Li and Yang}]{Wu2014}
\bibinfo{author}{Wu, P.}, \bibinfo{author}{Zhang, W.}, \bibinfo{author}{Li,
  Z.}, \bibinfo{author}{Yang, J.}, \bibinfo{year}{2014}.
\newblock \bibinfo{title}{Mechanisms of {{Graphene Growth}} on {{Metal
  Surfaces}}: Theoretical {{Perspectives}}}.
\newblock \bibinfo{journal}{Small} \bibinfo{volume}{10},
  \bibinfo{pages}{2136--2150}.
\newblock \DOIprefix\doi{10.1002/smll.201303680}.
\bibitem[{Yu et~al.(2008)Yu, Lian, Siriponglert, Li, Chen and Pei}]{Yu2008}
\bibinfo{author}{Yu, Q.}, \bibinfo{author}{Lian, J.},
  \bibinfo{author}{Siriponglert, S.}, \bibinfo{author}{Li, H.},
  \bibinfo{author}{Chen, Y.P.}, \bibinfo{author}{Pei, S.S.},
  \bibinfo{year}{2008}.
\newblock \bibinfo{title}{Graphene segregated on {{Ni}} surfaces and
  transferred to insulators}.
\newblock \bibinfo{journal}{Applied Physics Letters} \bibinfo{volume}{93},
  \bibinfo{pages}{113103}.
\newblock \DOIprefix\doi{10.1063/1.2982585}.
\bibitem[{Zandvliet(2006)}]{Zandvliet2006}
\bibinfo{author}{Zandvliet, H.J.W.}, \bibinfo{year}{2006}.
\newblock \bibinfo{title}{The {{2D Ising}} square lattice with nearest- and
  next-nearest-neighbor interactions}.
\newblock \bibinfo{journal}{Europhysics Letters (EPL)} \bibinfo{volume}{73},
  \bibinfo{pages}{747--751}.
\newblock \DOIprefix\doi{10.1209/epl/i2005-10451-1}.
\bibitem[{Zandvliet(2015)}]{Zandvliet2015}
\bibinfo{author}{Zandvliet, H.J.W.}, \bibinfo{year}{2015}.
\newblock \bibinfo{title}{Step free energy of an arbitrarily oriented step on a
  rectangular lattice with nearest-neighbor interactions}.
\newblock \bibinfo{journal}{Surface Science} \bibinfo{volume}{639},
  \bibinfo{pages}{L1--L4}.
\newblock \DOIprefix\doi{10.1016/j.susc.2015.04.015}.

\end{thebibliography}

\end{document}